\documentclass[sigconf, nonacm]{acmart}
\usepackage{algpseudocode}
\usepackage{algorithm}
\usepackage{tikz,tikzpeople}
\usepackage{amsmath}
\usetikzlibrary{shapes.geometric}
\usepackage{graphicx}
\usepackage{pgfplots}
\pgfplotsset{compat=1.5}
\usepackage{xcolor}
\usepackage[T1]{fontenc}
\usepackage[utf8]{inputenc}
\usepackage{multirow}
\usepgfplotslibrary{groupplots}

\newcommand\vldbdoi{10.14778/3485450.3485455}
\newcommand\vldbpages{31 - 45}
\newcommand\vldbvolume{15}
\newcommand\vldbissue{1}
\newcommand\vldbyear{2022}
\newcommand\vldbauthors{\authors}
\newcommand\vldbtitle{\shorttitle} 
\newcommand\vldbavailabilityurl{https://github.com/ArjitJ/DIAL}
\newcommand\vldbpagestyle{empty} 

\newcommand{\eat}[1]{}
\newcommand{\mysecref}[1]{Section \ref{#1}}

\newcommand{\myeqnref}[1]{Equation \ref{#1}}
\newcommand{\myfigref}[1]{{F}igure \ref{#1}}

\newcommand{\squishlist}{
\begin{list}{$\bullet$}
{ \setlength{\itemsep}{0pt} \setlength{\parsep}{3pt}
\setlength{\topsep}{3pt} \setlength{\partopsep}{0pt}
\setlength{\leftmargin}{0em} \setlength{\labelwidth}{1em}
\setlength{\labelsep}{0.5em} } }

\newcommand{\squishlisttwo}{
\begin{list}{$\bullet$}
{ \setlength{\itemsep}{0pt} \setlength{\parsep}{0pt}
\setlength{\topsep}{0pt} \setlength{\partopsep}{0pt}
\setlength{\leftmargin}{2em} \setlength{\labelwidth}{1.5em}
\setlength{\labelsep}{0.5em} } }

\newcommand{\squishend}{
\end{list}  }

\begin{document}
\newcommand{\sysname}{DIAL}
\newcommand{\cand}{\textsc{cand}}
\newcommand{\dups}{\textsc{dups}}
\newcommand{\train}{T}
\newcommand{\sel}{\textsc{sel}}
\newcommand{\test}{\textsc{test}}

\title{Deep Indexed Active Learning for Matching Heterogeneous Entity Representations}
%

\author{Arjit Jain}
\affiliation{%
  \institution{IIT Bombay}
}
\email{arjit@cse.iitb.ac.in}

\author{Sunita Sarawagi}
\affiliation{%
  \institution{IIT Bombay}
}
\email{sunita@iitb.ac.in}

\author{Prithviraj Sen}
\affiliation{%
  \institution{IBM Research, Almaden}
}
\email{senp@us.ibm.com}

\begin{abstract}

Given two large lists of records, the task in entity resolution (ER) is to find the pairs from the Cartesian product of the lists that correspond to the same real world entity. 
Typically, passive learning methods on such tasks require large amounts of labeled data to yield useful models. Active Learning is a promising approach for ER in low resource settings. However, the search space, to find informative samples for the user to label, grows quadratically for instance-pair tasks making active learning hard to scale. Previous works, in this setting, rely on hand-crafted predicates, pre-trained language model embeddings, or rule learning to prune away unlikely pairs from the Cartesian product. This blocking step can miss out on important regions in the product space
leading to low recall. We propose \sysname, a scalable active learning approach that jointly learns embeddings to maximize recall for blocking and accuracy for matching blocked pairs. 
\sysname\ uses an Index-By-Committee framework, where each committee member learns representations based on powerful pre-trained transformer language models.  We highlight surprising differences between the matcher and the blocker in the creation of the training data and the objective used to train their parameters.  Experiments on five benchmark datasets and a multilingual record matching dataset show 
the effectiveness of our approach in terms of precision, recall and running time.  
\end{abstract}

\maketitle
\pagestyle{\vldbpagestyle}
\begingroup\small\noindent\raggedright\textbf{PVLDB Reference Format:}\\
\vldbauthors. \vldbtitle. PVLDB, \vldbvolume(\vldbissue): \vldbpages, \vldbyear.\\
\href{https://doi.org/\vldbdoi}{doi:\vldbdoi}
\endgroup
\begingroup
\renewcommand\thefootnote{}\footnote{\noindent
This work is licensed under the Creative Commons BY-NC-ND 4.0 International License. Visit \url{https://creativecommons.org/licenses/by-nc-nd/4.0/} to view a copy of this license. For any use beyond those covered by this license, obtain permission by emailing \href{mailto:info@vldb.org}{info@vldb.org}. Copyright is held by the owner/author(s). Publication rights licensed to the VLDB Endowment. \\
\raggedright Proceedings of the VLDB Endowment, Vol. \vldbvolume, No. \vldbissue\ %
ISSN 2150-8097. \\
\href{https://doi.org/\vldbdoi}{doi:\vldbdoi} \\
}\addtocounter{footnote}{-1}\endgroup

\ifdefempty{\vldbavailabilityurl}{}{
\vspace{.3cm}
\begingroup\small\noindent\raggedright\textbf{PVLDB Artifact Availability:}\\
The source code, data, and/or other artifacts have been made available at \url{\vldbavailabilityurl}.
\endgroup
}

\section{Introduction}
\label{sec:introduction}

Entity resolution (ER) is a crucial task in data integration whose goal is to determine whether two mentions refer to the same real-world entity. With a history going back at least half a century (following \citet{FellegiSunter}'s seminal work), the task goes by various names and formulations, with the most common one being: Given two sets $R$ and $S$, for each pair of instances $(r,s) \in R \times S$ classify $(r,s)$ as either being a {\tt match} or a {\tt non-match}. In essence, this is an instance of \emph{paired classification} that requires learning a highly accurate binary-class classifier or \emph{matcher}.

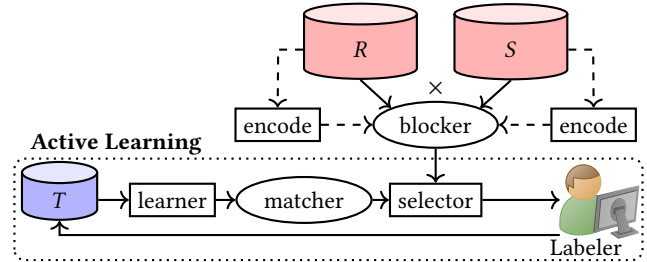
\begin{figure}
\begin{tikzpicture}
\node[cylinder,draw=black,thick,aspect=0.7,shape border rotate=90,minimum height=1cm,minimum width=1.5cm,cylinder uses custom fill, cylinder body fill=red!30,cylinder end fill=red!10] (r) at (0,0) {$R$};
\node[cylinder,draw=black,thick,aspect=0.7,shape border rotate=90,minimum height=1cm,minimum width=1.5cm,cylinder uses custom fill, cylinder body fill=red!30,cylinder end fill=red!10] (s) at (2,0) {$S$};
\node at (1,-0.5) {$\times$};
\node[ellipse,draw=black,thick] (block) at (1,-1) {blocker};
\node[rectangle,draw=black,thick] (encR) at (-1.1,-1) {encode};
\node[rectangle,draw=black,thick] (encS) at (3.1,-1) {encode};
\node[cylinder,draw=black,thick,aspect=0.7,shape border rotate=90,minimum height=0.7cm,minimum width=1cm,cylinder uses custom fill, cylinder body fill=blue!30,cylinder end fill=blue!10] (t) at (-4,-2) {$T$};
\node[rectangle,draw=black,thick] (learner) at (-2.5,-2) {learner};
\node[ellipse,draw=black,thick] (matcher) at (-0.75,-2) {matcher};
\node[rectangle,draw=black,thick] (selector) at (1,-2) {selector};
\node[person,monitor,minimum size=0.7cm] (person) at (3,-2) {Labeler};

\draw[->,thick] (r.south) -- (block.north west);
\draw[->,thick] (s.south) -- (block.north east);
\draw[->,dashed,thick] (r.west) -- +(-0.33,0) -- (encR.north);
\draw[->,dashed,thick] (s.east) -- +(0.33,0) -- (encS.north);
\draw[->,dashed,thick] (encR.east) -- (block.west);
\draw[->,dashed,thick] (encS.west) -- (block.east);
\draw[->,thick] (block) -- (selector);
\draw[->,thick] (t) -- (learner);
\draw[->,thick] (learner) -- (matcher);
\draw[->,thick] (matcher) -- (selector);
\draw[->,thick] (selector) -- (person.west);
\draw[->,thick] (person.south west) -- +(-6.65,0) -- (t.south);

\draw[draw=black,thick,dotted,rounded corners] (-4.6,-2.8) rectangle (3.75,-1.45);
\node at (-3.25,-1.25) {{\bf Active Learning}};
\end{tikzpicture}
\caption{Traditional paired classification AL.}
\label{fig:traditionalAL}
\end{figure}

ER has a rich history of employing active learning (AL) \citep{settles:uwtr09} instead of supervised or passive learning which harbors some advantages such as incrementally adding labeled pairs instead of requiring voluminous labeled data up-front to train the matcher. A variety of previous works on ER \citep{sarawagi02,10.14778/2735471.2735474,DBLP:conf/cikm/QianPS17,DBLP:conf/acl/KasaiQGLP19,DBLP:conf/sigmod/Meduri0SS20} have utilized the AL workflow shown in \myfigref{fig:traditionalAL}, with minor modifications. In each iteration, the learning algorithm (\emph{learner}) learns a \emph{matcher}  (shown in an ellipse which we use to denote model components) from $T$, the labeled pairs collected from the (human) labeler so far, while the example selector (\emph{selector}) chooses the most informative unlabeled pairs to acquire labels for. After including the new labels into $T$, the process repeats until we learn a matcher of sufficient quality. Popular choices for matcher includes support vector machines \citep{sarawagi02}, random forests \citep{DBLP:conf/sigmod/Meduri0SS20}, and neural networks \citep{DBLP:conf/acl/KasaiQGLP19}. Popular choices for selector includes query-by-committee \citep{seung:colt92,freund:ml97} which has seen wide usage in ER \citep{10.14778/2735471.2735474,10.1007/978-3-030-28730-6_5} and uncertainty sampling~\citep{DBLP:conf/acl/KasaiQGLP19}.

To efficiently pare down the number of unlabeled pairs in $R \times S$ that the selector needs to choose from, one usually employs a pre-specified \emph{blocking} function. Commonly used blocking functions include string similarity measures (e.g. Jaccard similarity) to compare string representations of $r$ and $s$, and keep only those pairs whose similarity exceeds a pre-determined threshold \citep{DBLP:conf/sigmod/Meduri0SS20}. \citet{10.14778/2994509.2994535} recommend that the user acquire some domain knowledge about $R,S$ so as to be able to specify an effective blocker. Even if domain knowledge is available, the user's choice may still be suboptimal. In some situations, it may even be impossible to acquire such knowledge, for example when one of $R$ or $S$ is in a language unfamiliar to the user (aka cross-lingual ER \citep{mcnamee-etal-2011-cross}). The other, possibly more disconcerting, conceptual issue with \myfigref{fig:traditionalAL} is that the blocker is removed from the matcher. To be clear, both matcher and blocker are paired classifiers but the requirements of them are different. While the matcher needs to provide high classification accuracy, the blocker only needs to efficiently identify matches while rejecting as many non-matches as possible (in other words, high recall is desired). This implies that ideally, the blocker should be integrated into the AL feedback loop. As we obtain more labeled data, we expect both matcher and blocker to benefit instead of benefiting one and not the other as \myfigref{fig:traditionalAL} indicates. While there exist proposals to learn the blocking function automatically \citep{bilenko2006adaptive}, these require copious amounts of labeled data up-front and thus it is not clear how to combine this with low-resource AL setting where such labeled data may not be available. Optionally, DeepER \citep{10.14778/3236187.3236198} encodes $r \in R$ and $s \in S$ into fixed-dimensional vectors or \emph{encodings} $E(r)$ and $E(s)$, respectively, (see dashed edges in \myfigref{fig:traditionalAL}). Similar pairs $(r,s)$ may then be retrieved via nearest neighbor search implemented using locally sensitive hashing. However, even in this case the blocker is not integrated with the AL feedback loop.

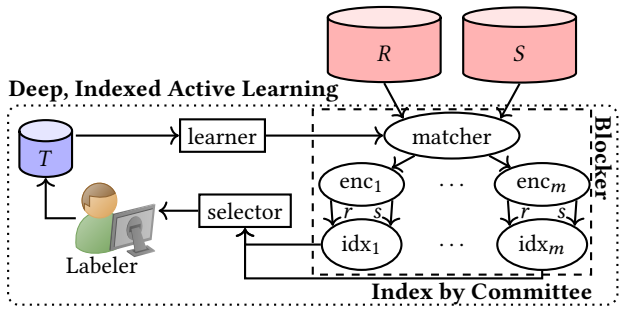
\begin{figure}
\begin{tikzpicture}
\node[cylinder,draw=black,thick,aspect=0.7,shape border rotate=90,minimum height=1cm,minimum width=1.5cm,cylinder uses custom fill, cylinder body fill=red!30,cylinder end fill=red!10] (r) at (4.5,1.2) {$R$};
\node[cylinder,draw=black,thick,aspect=0.7,shape border rotate=90,minimum height=1cm,minimum width=1.5cm,cylinder uses custom fill, cylinder body fill=red!30,cylinder end fill=red!10] (s) at (6.3,1.2) {$S$};
\node[cylinder,draw=black,thick,aspect=0.7,shape border rotate=90,minimum height=0.7cm,minimum width=0.7cm,cylinder uses custom fill, cylinder body fill=blue!30,cylinder end fill=blue!10] (t) at (0,-0.2) {$T$};
\node[rectangle,draw=black,thick] (learner) at (2.35,0.1) {learner};
\node[ellipse,draw=black,thick] (matcher) at (5.4,0.1) {matcher};
\node[ellipse,draw=black,thick,minimum width=0.9cm] (enc1) at (4.2,-0.55) {$\text{enc}_1$};
\node at (5.4,-0.55) {$\ldots$};
\node[ellipse,draw=black,thick,minimum width=0.9cm] (encm) at (6.6,-0.55) {$\text{enc}_m$};
\node[ellipse,draw=black,thick,minimum width=0.9cm] (idx1) at (4.2,-1.35) {$\text{idx}_1$};
\node at (5.4,-1.35) {$\ldots$};
\node[ellipse,draw=black,thick,minimum width=0.9cm] (idxm) at (6.6,-1.35) {$\text{idx}_m$};
\node[rectangle,draw=black,thick] (selector) at (2.65,-0.9) {selector};
\node[person,monitor,minimum size=0.7cm] (person) at (0.75,-1) {Labeler};

\draw[->,thick] (r.south) -- (matcher.north west);
\draw[->,thick] (s.south) -- (matcher.north east);
\draw[->,thick] (0.4,0.1) -- (learner);
\draw[->,thick] (learner) -- (matcher);
\draw[->,thick] (matcher) -- (enc1);
\draw[->,thick] (matcher) -- (encm);
\draw[->,thick] (enc1.south west) -- (idx1.north west) node[midway, right, xshift=1pt] {$r$};
\draw[->,thick] (enc1.south east) -- (idx1.north east) node[midway, left] {$s$};
\draw[->,thick] (encm.south west) -- (idxm.north west) node[midway, right, xshift=1pt] {$r$};
\draw[->,thick] (encm.south east) -- (idxm.north east) node[midway, left] {$s$};
\draw[->,thick] (idx1.west) -- (2.65,-1.35) -- (selector.south);
\draw[->,thick] (idxm.south) -- (6.6,-1.8) -- (2.65,-1.8) -- (selector.south);
\draw[->,thick] (selector) -- (1.5,-0.9);
\draw[->,thick] (person.west) -- (0,-1) -- (t);

\draw[draw=black,thick,dotted,rounded corners] (-0.5,-2.15) rectangle (7.6,0.5);
\node at (1.7,0.7) {{\bf Deep, Indexed Active Learning}};
\draw[draw=black,thick,dashed] (3.55,-1.75) rectangle (7.25,0.45);
\node at (5.8,-2) {{\bf Index by Committee}};
\node[rotate=270] at (7.42,-0.2) {{\bf Blocker}};
\end{tikzpicture}
\caption{Deep, indexed AL with a committee of encoders.}
\label{fig:DIAL}
\end{figure}

Given the previous discussion, our goal is to propose a new AL approach for ER that satisfies the following desiderata:
\squishlisttwo
\item To ensure that both benefit from newly acquired labels, the blocker and matcher should be integrated.
\item Slightly at logger heads with the previous property, we would also like to train the matcher and blocker with distinct loss functions since they need to satisfy distinct requirements.
\item Need to achieve all this without adversely affecting scalability, since this is one of the purposes of the blocker in \myfigref{fig:traditionalAL}.
\squishend
Our proposed integrated matcher-blocker combination and new AL workflow is shown in \myfigref{fig:DIAL}. Compared to \myfigref{fig:traditionalAL}, the two most notable differences are 1) the blocker (dashed box) is now part of the AL feedback loop, and 2) the matcher is a component within the blocker. As base matcher, we use \emph{transformer-based pre-trained language models} (TPLM) \citep{devlin-etal-2019-bert,liu2019roberta} which have recently led to excellent ER accuracies in the passive (non-AL) setting \citep{ditto}. Before we describe details of the proposed approach, please note that TPLMs can be invoked in two distinct modes. To obtain a prediction for a pair $(r,s)$, we invoke the matcher, a (fine-tuned) instance of TPLM, in \emph{paired} mode where we concatenate string representations of $r$ and $s$ to obtain a joint representation. \emph{Single} mode is where we input one of $r$ or $s$'s string representation to obtain its encoding. To implement blocking, we invoke TPLM in single mode by first populating an index structure (FAISS \citep{faiss}) with $E(r), ~ \forall r \in R$, followed by probing with $E(s)$ given $s \in S$ to retrieve potentially similar pairs for the selector to choose from. Since the blocker needs to attain high recall, we find that one encoding of $r$ or $s$ is not sufficient, improved recall can be achieved if we allow minor variations of $E(r)$ and  $E(s)$. To this end, we allow for multiple, distinct affine transformations of the base TPLM's last layer. This combination of TPLM, multiple encoders $\text{{\tt enc}}_1, \ldots \text{{\tt enc}}_m$ and indices $\text{{\tt idx}}_1, \ldots \text{{\tt idx}}_m$, referred to as \emph{index by committee} (IBC), is pictorially depicted in \myfigref{fig:DIAL}. While previous works on AL and ER have attempted to learn committees instead of a single classifier \citep{10.1007/978-3-030-28730-6_5,sarawagi02,10.14778/2735471.2735474,DBLP:conf/sigmod/Meduri0SS20}, none of them consider TPLMs, and none of them use committee for designing a high recall blocker.

A primary challenge of our integrated matcher-blocker system is training them simultaneously so that the blocker recalls all likely duplicates based on embedding similarity, and the matcher precisely separates duplicates from non-duplicates in paired mode.  We show that the conventional classification loss on labeled examples that trains the matcher well, performs very poorly on the blocker.  This led us to design a new contrastive training objective for the blocker that separates labeled duplicates from {\em random} non-duplicate pairs.   In entity resolution tasks, random non-duplicates are generally much easier to separate out than the difficult non-duplicates selected during active learning~\cite{sarawagi02}.  While such difficult near-duplicates are essential for training a {\em precise matcher}, they interfere with the goals of co-embeddings duplicates during blocking. We observe dramatic drops in recall when blocker embeddings are trained with actively labeled non-duplicates, likewise dramatic drop in precision when the matcher's classifier is trained with random negatives.  These findings were key to our jointly training the matcher and blocker in an active learning loop so as to match or even surpass the yield obtained by hand-designed rules on existing benchmarks. Additionally, on heterogeneous entity lists where existing methods relied on pre-trained embeddings, we obtained significantly higher recall with our method of training the blocker.

\noindent Our contributions are:
\squishlisttwo
\item To the best of our knowledge, ours is the first active learning ER proposal to \emph{integrate} the matcher with the blocker.   With availability of more labels, improvements in the former directly benefits the latter. 

\item We design a novel method of learning record embeddings for the blocker using (1) a contrastive training objective and (2) random non-duplicates. This design choice is crucial to achieve high recall; and   provides as much as 25 points increase compared to using matcher embeddings as-is.  

\item Learning a committee of encodings is in itself a novel contribution. To the best of our knowledge, we are not aware of any previous work that can learn a committee of TPLMs. By combining with indexing, leads to IBC, a novel, fast and effective example blocking technique.
\item We evaluate the efficacy of our learned blocker by comparing with (1) hand-crafted blocking functions used in popular ER datasets, and (2) state of the art learned embeddings methods on a  multilingual dataset and five ER datasets.
\item \sysname\ provides an \emph{absolute} improvement on the F1 scores by $6-20\%$ on two product datasets, $4-10\%$ on a bibliographic dataset, $40-55\%$ on a textual dataset, and $5-18\%$ on a multilingual dataset, over baseline approaches demonstrating the effectiveness of \sysname\ across various real world datasets. On some of these datasets \sysname~produces even better recall than hand-tuned blocking functions  without any external knowledge about the domain and with only limited number of judiciously chosen label.    

\squishend
\eat{
\paragraph{Contributions}
\begin{itemize}
    \item First method to be able to train a committee using pretrained language models
    \item New method \sysname\ for simultaneously learning blocking predicates and matching functions via active learning. 
    Particularly useful when matching heterogeneous entity representations where dataset-specific rule-based blocking predicates are not available.
    \item \sysname\ is significantly more accurate than blocking on similarity of trained embeddings, and close to human provided rule-based blocking functions.
    \item  Committee used to create multiple embeddings has interesting differences from existing committees used in AL (Sampled negatives better than labeled negatives, orthogonality loss, Classifier on embeddings while useful for learning blocking predicates is not useful for AL where contrastive training loss is more useful)
    \item Compared many different methods of AL including SOTA methods like BADGE.
\end{itemize}
}
\eat{Limitations of existing method
\begin{itemize}
    \item Does not use committee
    \item Hand crafted blocking not available for heterogeneous representations
    \item Pretrained embeddings have poor recall
    \item All pairs has scalability issues
    \item Blocking is not part of the training loop
\end{itemize}}

\section{Preliminaries and Background}
We formally state our problem and provide relevant background in this section.
\subsection{Problem Statement}
\label{sec:problemstatement}
Given two large lists $R$ and $S$ of entities, our goal is to design an end to end system that can identify the subset $\dups$ of $R \times S$ that are duplicates across the two lists. $R$ and $S$ could be the same list, and the matchings could be many to many.  Each entry $r \in R$ or $s \in S$ could consist of one or more attributes that are predominantly textual. In general, the attributes across the lists may not be aligned, and the space of their values may be incomparable.  For example, list $R$ may list product names and descriptions in German whereas list $S$ may be in English. 
%
%
Our goal is to learn in an integrated active learning loop (1) a {\em blocker} to efficiently identify the subset $\cand$  of $R \times S$ that are likely duplicates, and (2) a {\em matcher} to assign a final verdict of duplicate or not for each entity pair $(r,s)$ in the filtered set $\cand$.  
We are given three types of resources:
a transformer based pretrained language model (TPLM),
%
a small seed labeled dataset $\train$ of duplicates and non-duplicate pairs, and 
a labeling budget $B$  of getting human labels on pairs selected from $R \times S$ to augment $\train$. 

\subsection{Pre-trained Language Models}
\label{sec:pretrainedlm}
Transformer based pretrained LMs (TPLM) such as BERT~\cite{devlin-etal-2019-bert} and RoBERTa~\cite{liu2019roberta} have been shown to transfer remarkably well to many different tasks and domains.  The input to the transformer is a sequence of tokens.  The transformer uses multiple layers of self-attention to output for each token a fixed dimensional contextual embedding. The hundreds of million parameters used in a transformer are pre-trained using large amounts of unlabeled text corpus e.g. Wikipedia. This results in assigning each word an embedding that captures its semantics in the context of the current sentence.  These highly contextual embeddings have been found useful in a number of downstream NLP tasks. In ER they have been shown to lead to robustness to spelling mistakes, and abbreviations, and provide state of the art performance on "dirty" datasets~\cite{ditto,DBLP:conf/edbt/BrunnerS20}. 
A standard approach to use these models in a new task is to add task specific layers on top of the transformer and fine-tune using a task specific objective. 
%
%
%
%
There are two common modes to fine-tune a transformer for a pairwise classification task required in ER.

\subsubsection{Paired mode}
\label{sec:paired}
In this mode the transformer is fed a concatenation of the tokens of the two records as follows:
    \begin{equation}
        [\text{CLS}],  r_1\ldots r_n, [\text{SEP}], s_1\ldots s_m, [\text{SEP}]
    \end{equation}
where $r_1,\ldots r_n$ denote tokens of record $r$, $s_1\ldots s_m$ denote tokens of $s$, CLS denotes a special start token and SEP denotes a special separator token.
The last layer of the transformer assigns fixed $d$ dimensional contextual embeddings to all $m+n+3$ tokens.
The contextual embedding  of the $[\text{CLS}]$ token is treated as an embedding $E(r,s)$ of the pair. This embedding is used to classify the pair as duplicates or not via additional light-weight layers.  This is the mode we use for the matcher since the learned attention across tokens in the records can focus on distinguishing words. %
Consider an example of a pair of records describing two different editions of the same book. An embedding based model can have a hard time trying to distinguish these two instances, however, a transformer model can by aligning the attention between the tokens corresponding to book edition between the two instances. Other examples include the price attribute in a products dataset, and house number in a postal addresses dataset.

\subsubsection{Single mode}
\label{sec:single}
The above paired mode is not practical to invoke on every $(r,s)$ in the Cartesian product $R \times S$. A second way is to first separately encode each record. For a record $x$ in $R$ or $S$ we obtain its embedding from the TPLM by first feeding to the transformer:
\begin{equation}
        [\text{CLS}],  x_1\ldots x_n [SEP]
\end{equation}
where $x_1,\ldots x_n$ denote the tokens in record $x$.  We obtain fixed $d$ dimensional contextual embeddings $E(x_1),\ldots E(x_n)$ from the TPLM. We then define the embedding of the record $x$ as the mean of its token embeddings. 
\begin{equation}
\label{eq:embed}
    E(x) = \frac{1}{n} \sum_{i=1}^n E(x_i)
\end{equation}
For a pair of records $(r,s)$ we separately compute embeddings $E(r)$ and $E(s)$ and decide on whether they are duplicate or not based only on these fixed embeddings.
A well-known example is
SentenceBERT \cite{sentencebert} whose classifier 
%
takes as input the concatenation of the embedding $E(r)$ of $r$, embedding $E(s)$ of $s$, and the absolute element-wise difference between the two embeddings $|E(r)-S(s)|$ and adds a linear layer above it. After training with appropriate labeled data, these embeddings can be used for efficient nearest neighbour search to retrieve likely duplicates. We will harness the single mode for the design of our blocker.




\subsection{Example selection for ER}
\label{sec:exampleselection}

\subsubsection{Query-by-Committee via Bootstrap}
\label{sec:qbc}
Query-by-Committee (QBC) \citep{seung:colt92,freund:ml97} has a rich history of application in ER going back to ALIAS \citep{sarawagi02}. We review 1) the bootstrap-based classifier-agnostic approach towards building a committee \citep{10.14778/2735471.2735474}, followed by 2) selecting examples for labeling using said committee. While there exist many techniques to build a committee of classifiers given the same labeled data, most of these are specifically designed for certain classifiers, e.g., randomizing the choice of the feature to split on while adding a node in the decision tree is a specific technique to learn a committee of decision trees \citep{sarawagi02}. \citeauthor{10.14778/2735471.2735474} propose \emph{bootstrap} as a way to build a committee that is \emph{agnostic} to the classifier being used. Given labeled data $T$, bootstrap creates multiple versions $T_1, \ldots T_m$ by sampling from $T$ with replacement so that each $T_i$ contains the same number of pairs as $T$. Subsequently, one may use $T_i$ to train a member of the committee by using it as training data. Given an unlabeled pair $(r,s)$, one may then compute the variance in its predicted label as: \begin{equation*}
    var(r,s) = \frac{\text{{\tt \#match}}(r,s)}{m} \left(1 - \frac{\text{{\tt \#match}}(r,s)}{m}\right)
\end{equation*}
where {\tt \#match}$(r,s)$ denotes the number of committee members predicting $(r,s)$ to be a duplicate out of the $m$-sized committee. Pairs with higher variance are selected for labeling.

\subsubsection{Uncertainty Sampling}
Besides variance, other metrics are also available to measure the uncertainty of the prediction for $(r,s)$. These may be used independent of the committee, especially when the classifier produces prediction probabilities besides the label. DTAL \citep{DBLP:conf/acl/KasaiQGLP19} uses (conditional) \emph{entropy}:
\begin{equation}
\label{eq:uncertain}
    H(p) =~ -p \log p - (1-p)\log(1-p)
\end{equation}
where $p$ denotes $\Pr(y=\text{{\tt match}} | (r,s))$, the predicted probability of $(r,s)$ being a match.

\subsubsection{High Confidence Sampling with Partition} 
Besides entropy, DTAL~\citep{DBLP:conf/acl/KasaiQGLP19} also proposes High Confidence Sampling with Partition. They divide the candidate set into two subsets consisting of pairs that are predicted as positives, and negatives respectively by the matcher. From both these sets they choose an equal amount of most confident and least confident pairs, based on their entropy, giving four sets, $p_{hc}, p_{lc}, n_{hc}, n_{lc}$ representing high and low confidence positives, and high and low confidence negatives respectively. They query the user to label $p_{lc}$ and $n_{lc}$, but they do NOT query the user to label $p_{hc}$ and $n_{hc}$. Instead, they directly add them to the labeled positives and negatives, i.e.
$ \train_p \gets \train_p \cup p_{hc}$ and $ \train_n \gets \train_n \cup n_{hc}$
\subsubsection{BADGE} In a batch active learning setup, BADGE \cite{badge} tries to combine uncertainty and diversity for example selection by computing hallucinated gradient embeddings. Given a neural network classifier $f(x; \theta)$, with weights $\theta_0$, and a query point $x$ from the candidate set, BADGE calculates $\hat{y}$, the most likely label for $x$ according to the class probabilities output by $f$. It then uses $\hat{y}$ to compute the gradient embedding 
$$
g_{x}=\left.\frac{\partial}{\partial \theta_{\text {out }}} \ell(f(x ; \theta), \hat{y})\right|_{\theta=\theta_{0}}
$$
where $\theta_{\text{out}}$ refers to the parameters of the output layer, and $\ell$ is a loss function, usually taken to be the standard cross entropy loss. Notice that the magnitude of these gradient embeddings can be used as a proxy for uncertainty, as confident samples will have lower gradient magnitudes. 
To incorporate diversity, examples to query the user are selected using the k-\textsc{means}++ \cite{kmeanspp} seeding algorithm on the set $\{g_x: x \in \cand\}$. 

\section{\sysname}
\sysname\ starts with an initial set of labeled pairs $\train$ of duplicates and non-duplicates and iteratively collects $B$ more labeled pairs in an active learning loop.  In each iteration of the loop, it performs the following steps: (1) trains a Matcher model that given a pair of records can assign a probability of the pair being duplicate, (2) trains a Blocker model to encode records in $R$ or $S$ so that duplicates are close, (3) performs an indexed nearest neighbor search over the encodings to filter a candidate set $\cand \subset R\times S$  of likely duplicate pairs, (4) selects a subset $\sel$ of $\cand$ using uncertainty assignments from Matcher, (5) collects user's duplicate or not labels on pairs in $\sel$ and augments $\train$.  
At the end of the loop, all pairs in the candidate set predicted duplicates by the Matcher are returned as the duplicate set. 
This labeling loop differs from earlier AL-based ER systems in one crucial way.  While existing systems assume a fixed candidate set $\cand$ under a user-provided or pre-trained blocking function, we propose to learn a Blocker and adaptively create candidates $\cand$ within the AL loop.  Our challenge then is how to perform this step while ensuring that our learned Blocker can match hand-crafted rules in terms of recall, and do that without enumerating the Cartesian product $R \times S$.

\begin{figure*}
    \centering
    \includegraphics[width=\linewidth]{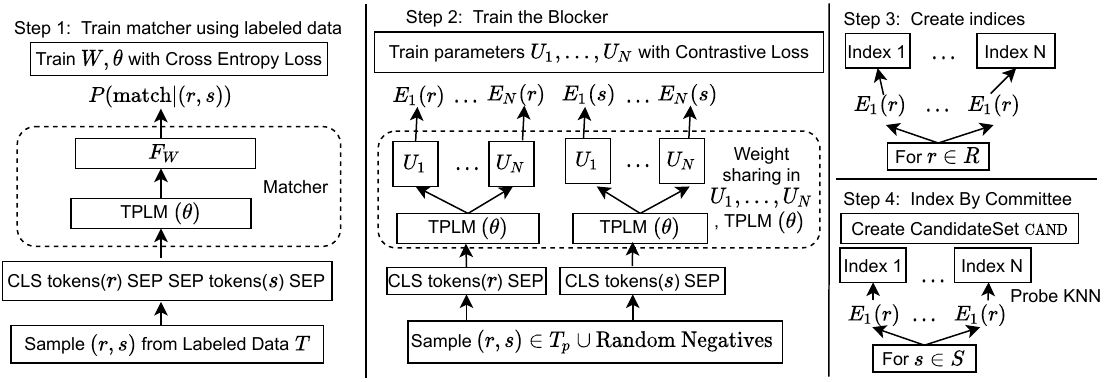}
    \caption{Outline of the proposed system. \sysname\ integrates TPLM based matcher and blocker models. In each iteration of the active learning loop, it performs the following steps: trains a Matcher model that given a pair of records can assign a probability of the pair being duplicate, trains a Blocker model to independently encode records in $R$ or $S$, and performs an indexed nearest neighbor search over the encodings to filter a candidate set $\cand \in R \times S$ of likely duplicate pairs. Candidate set $\cand$ is used by the selector to obtain a subset of samples to be labeled by the user.}
    \label{fig:dial}
\end{figure*}
Our matcher and blocker are integrated and both leverage TPLMs. 
We present the design of the main modules of \sysname. An overview appears in Figure~\ref{fig:dial}.

\subsection{Matcher}
\label{sec:matcher}
For each record pair $(r,s)$ the matcher needs to assign a probability $\Pr(y=1|(r,s))$ of the pair being a duplicate. The matcher uses the transformer in the paired mode described in Section~\ref{sec:paired} to get a joint embedding $E(r,s) \in R^d$ of $(r,s)$. Let $\Theta$ denote all the parameters of the transformer. These embeddings are converted into a probability of the pair being duplicates using additional neural layers $F_W:R^d\mapsto R$:
\begin{equation}
    \Pr(y=1|(r,s)) = (1+\exp(-F_W(E(r,s)))^{-1}
\end{equation}
where $W$ denote parameters of the matcher specific layers to be learned along with parameters $\Theta$ of the transformer.  In our case, $F_W$ comprised of a linear layer, followed by a tanh activation, followed by another linear layer to get a single scalar score which is then converted into a probability using the above sigmoid function.
During training, the initial values of parameters $\Theta$ are from the TPLM whereas $W,b$ take random values. All  three sets of parameters are optimized using the standard cross entropy loss on the labeled training set $T$.
\begin{equation}
\label{eq:matcher}
\begin{split}
    \min_{\Theta, W} \sum_{(r^i,s^i)\in \train_p} \log(1+\exp(-F_W(E_\Theta(r^i,s^i)))) \\ + \sum_{(r^i,s^i)\in \train_n} \log(1+\exp(F_W(E_\Theta(r^i,s^i))))
    \end{split}
\end{equation}
where $\train_p$ denotes the duplicate pairs in $\train$ and $\train_n = \train-\train_p$ denotes the non-duplicates.  In the above we put the subscript $\Theta$ on the embeddings to denote that the transformer parameters are further fine-tuned to achieve the matcher's goal of assigning probability close to 1 to the duplicates and close to 0 to the non-duplicates. See Step~1 in Figure~\ref{fig:dial} for a summary of this part.
\newcommand{\randr}{\text{rand}(R)}
\newcommand{\rands}{\text{rand}(S)}
\newcommand{\numC}{N}
\subsection{Blocker}
\label{sec:blocker}
Here our goal is to obtain embeddings of each record in $R$ and $S$ so we can retrieve likely duplicates via nearest neighbor search.
Existing methods for getting such embeddings is to use the transformer in {\em single} mode as outlined in Section~\ref{sec:single}, either as-is or with further fine-tuning using $\train$ as in SentenceBERT~\cite{sentencebert,ditto}.  We will show in Section~\ref{sec:expBlocker} that both methods perform surprisingly poorly in retrieving duplicates.  
The blocker in \sysname\ makes three important design choices that jointly provide significant gains over existing methods.  We outline each of these next. 

\subsubsection{Index by Committee of Embeddings (IBC)}
\label{sec:committee}
Our blocker assigns a committee of $\numC$ different embeddings to a record in $R$ or $S$.
Traditionally in AL, committees (Section~\ref{sec:qbc}) are used to assign uncertainty values during example selection.  Here, we propose to use multiple embeddings for a different goal of casting a wider net so that all likely duplicates are covered in {\em any one} of the $\numC$ embeddings.

We start with the $d$-dimensional embeddings $E(x)$ obtained from the Matcher-trained Transformer operating in single mode as described in Equation~\ref{eq:embed}.  Then we create a committee of $\numC$ different light-weight layers to produce  a set of $\numC$ $d$-dimensional embeddings: $E_1,\ldots E_{\numC}$.
Each committee member $k$, first chooses a fixed random mask $M_k \in \{0,1\}^d$ to retain only a random fraction $p$ of the initial embeddings $E(x)$.  This step is inspired by the choice of random attribute selection in random forests~\cite{randomForests}. Then a linear layer transforms the masked embeddings via learned parameters to obtain the $k$-th embedding vector $E_k(x)$ as:
\begin{equation}
    \label{eq:committeeembed}
    E_k(x) = \tanh(U_k (M_k \odot E(x), 1))
\end{equation}
where $U_k \in R^{d(d+1)}$ denote the learned parameters used to obtain the $k$-th embedding vector of record $x$.  The transformer parameters $\Theta$ used to compute $E(x)$ are not trained by the blocker.  We next describe how we train the $U_k$ parameters. 

\subsubsection{Choice of Training data}
\label{sec:blockertrainingdata}
One subtle problem we encountered is using the labeled data $\train$ collected via AL to train the blocker. The negatives in $\train$ are mostly {\em near-duplicates} and were chosen by AL because they were hard to separate from duplicates.  While such hard negatives are extremely useful for learning a precise matcher as several previous AL work have shown~\cite{sarawagi02,DBLP:conf/acl/KasaiQGLP19}, they are detrimental to learning good embeddings for blocking where the goal is high recall rather than high-precision.  Embeddings trained to separate the similar non-duplicates $\train_n$ from the actual duplicates $\train_p$, might also throw the unseen duplicates apart.   We therefore create easier non-duplicates in the following way:

Given a set $D_p$ of $b$ duplicates in a training batch, we randomly sample a set $\randr$ of $b$ records from $R$ and an independent random set $\rands$ of $b$ records from $S$.
We then obtain embeddings $E(x)$ for all records in \randr,\rands, and each record in $D_p$ the duplicate pairs.  Now each committee randomly shuffles the set of records in $\randr,\rands$ and obtains a random set of $b$ non-duplicate pairs $(r_1,s_1)\ldots (r_b,s_b)$ by concatenating the shuffled lists.  Further, for   
each duplicate pair $(r_p,s_p)$ in the training batch  $D_p$ we obtain further non-duplicates as $(r_p, s_i), (r_i, s_p)$ for $i=1\ldots b$.  

\subsubsection{Choice of Training Objective}
Given the set of duplicates and non-duplicates, a default training objective would be impose a binary classification loss to separate them as is done for the matcher in Eq~\ref{eq:matcher}. However, again considering the differing goals of the two systems, we propose a different contrastive training objective that jointly separates a duplicate from all non-duplicates.  The contrastive loss requires a similarity function $\text{sim}(u,v)$ between any two embedding vectors $u,v$. The training objective of the $k$-th committee member is then 
\begin{align}
\label{eq:contrast}
  &\max_{U_k}\sum_{(r_p,s_p)\in T_p}\log\left[\dfrac{{\text{s}(r_p, s_p)}}{{\text{s}(r_p, s_p)} + \sum\limits_{i=1}^b \left({\text{s}(r_i, s_p)} + {\text{s}(r_p, s_i)} + {\text{s}(r_i, s_i)}\right)}\right] \\
  & \text{where} ~~~~ s(r,s) = e^{\text{sim}(E_k(r), E_k(s))} \nonumber
\end{align}

We use the negative squared $\ell_2$ distance as a similarity function. Scaled cosine similarity is another good choice.  The only requirement is that we should be able to retrieve nearest neighbour efficiently using that similarity function.

Step~2 in Figure~\ref{fig:dial} summarizes the training of the blocker.  Notice the differences in the input, the training objective, and the training dataset with the training of the matcher in step~1.


\subsection{Overall Algorithm}
Algorithm \ref{alg:dial} outlines the pseudo-code of \sysname. In each round of Active Learning, \sysname\ first trains the TPLM parameters $\Theta$, and parameters $W$ of the matcher specific layer $F_W$ with the binary classification objective (Equation~\ref{eq:matcher}) on the labeled data $\train$. It then freezes the weights of parameters $\Theta$, and creates a committee where each member implements an embedding layer as described in Section \ref{sec:blocker} (Equation~\ref{eq:committeeembed}). To train the committee, it samples duplicate pairs from the labeled data $T_p$, creates random negative pairs $(r, s)$ where $r \in \randr$ and $s \in \rands$, and individually obtains the transformer representations for each of these. Every committee member computes individual embeddings for each of these instances, and is trained using the contrastive objective (Equation (\ref{eq:contrast})). After training the committee, each member creates an index on the embeddings of instances in $R$, and queries this index to get the $k$ nearest neighbours for each instance in $S$. The closest pairs across all members are used to construct the set $\cand$, which are fed to an active learning instance selector to select the most informative $B$ pairs to be labeled by the user.  \sysname\ is agnostic to the specific selection algorithm and we present results with many existing selection algorithms in Section~\ref{sec:selection}. Our default is uncertainty sampling (Eq~\ref{eq:uncertain}).
Figure \ref{fig:dial} highlights the main operations performed by \sysname\ in an active learning round, and clearly describes the data flow. 
\begin{algorithm}
\begin{algorithmic}[1]
\Require{TPLM with parameters $\Theta$, Lists $R$ and $S$, Seed Labeled Data $\train$, $\cand$ Size, Labeling Budget per round $B$,  Committee Size $\numC$, Number of neighbours $k$}
\ForAll{round of Active Learning}
    \State {\bf $\triangleright$ Train the matcher}
	\State Find ${\Theta, W}$ that minimize Eq. (\ref{eq:matcher}) using $T$
	\State {\bf $\triangleright$ Create committee: each member $k$, has trainable parameters $U_k$, computes embedding $E_k(x)$ using Eq. (\ref{eq:committeeembed})}
    \State {\bf $\triangleright$ Train the embeddings}
    \ForAll{committee member $k$}
		\State Find ${U_k}$ that maximize Eq. (\ref{eq:contrast}) using $T_p\ \&$ Random Negatives (See Sec \ref{sec:blockertrainingdata})
	\EndFor
	\State {\bf $\triangleright$ Retrieving Pairs}
    \State Create Indexes $\text{IDX}_i$ for each committee member $i$
	\ForAll{$r$ in $R$}
	    \State Compute TPLM embedding $E(r)$
	    \ForAll{committee member $k$}
	        \State Add $E_k(r)$ to $\text{IDX}_k$
	    \EndFor
	\EndFor
	\State Create list RP to store Retrieved Pairs 
	\State RP = []
	\ForAll{$s$ in $S$}
	    \State Compute TPLM embedding $E(s)$
	    \ForAll{committee member $c$}
	        \State Add $k$ nearest neighbours of $E_c(s)$ in $\text{IDX}_c$ to RP
	    \EndFor
	\EndFor
	\State Create $\cand$ containing the closest pairs from RP
	\State Select $B$ pairs from $\cand$ $\&$ query user labels. (See Sec~\ref{sec:selection}) 
	\State Update $\train$ with the newly labeled data
\EndFor
\end{algorithmic}
\caption{Pseudo-Code of the proposed system \sysname.}
\label{alg:dial}
\end{algorithm}


\newcommand{\pairedFixed}{PairedFixed}
\newcommand{\onlyPaired}{PairedAdapt}
\newcommand{\ditto}{DittoAB}
\newcommand{\dittoAdapt}{SentenceBERT}
\newcommand{\human}{Rules}
\newcommand{\walmartamazon}{Walmart-Amazon}
\newcommand{\amazongoogle}{Amazon-Google}
\newcommand{\dblpacm}{DBLP-ACM}
\newcommand{\dblpscholar}{DBLP-Scholar}
\newcommand{\abtbuy}{Abt-Buy}
\newcommand{\multilingual}{MultiLingual}

\section{Experiments}
We present an extensive comparison of \sysname\ with existing methods based on hand-crafted predicates, learned embeddings, and existing meta-blocking methods.  We also present a detailed ablation study and analyze \sysname's running time. 
\subsection{Datasets}
\label{sec:datasets}
\begin{table}[t]
    \centering
    \caption{Statistics reporting the scale of the datasets used to evaluate \sysname.}
    \begin{tabular}{l|c|c|c|c|c}
    \toprule
         Dataset &  $|R|$ &  $|S|$ &  $|\dups|$ &  $|\frac{\dups}{R \times S}|$ &  $|\mathcal{D}_{\text{test}}|$ \\
    \midrule
         \walmartamazon & $2554$ & $22074$ & $1154$ & $\sim 2\mathrm{e}{-5}$ & $2049$\\
         \amazongoogle & $1363$ & $3226$ & $1300$ & $\sim 3\mathrm{e}{-4}$ & $2293$\\
         \dblpacm & $2616$ & $2294$ & $2224$ & $\sim 3\mathrm{e}{-4}$ & $2473$\\
         \dblpscholar & $2616$ & $64263$ & $5347$ & $\sim 3\mathrm{e}{-5}$ & $5742$\\
         \abtbuy & $1081$ & $1092$ & $1097$ & $\sim 1\mathrm{e}{-3}$ & $1916$\\
         \multilingual & $100\mathrm{k}$ & $100\mathrm{k}$ & $100\mathrm{k}$ & $\sim 1\mathrm{e}{-5}$ & $2000$\\
    \bottomrule
    \end{tabular}
    \label{tab:datasets}
\end{table}

We validate our approach on five widely used real world datasets from DeepMatcher~\cite{deepmatcher},
ER Benchmark \cite{10.14778/1920841.1920904} and the Magellan data repository \cite{magellandata} as summarized in Table~\ref{tab:datasets}. \walmartamazon, \amazongoogle\ and \abtbuy\ are product datasets, whereas \dblpacm\ and \dblpscholar\ are citation datasets. \abtbuy\ is a textual dataset, whereas the other four are structured datasets. To use \walmartamazon\ and \amazongoogle\ as structured datasets, we follow the schema used by DeepMatcher \cite{deepmatcher}. 
As we motivated in Section \ref{sec:problemstatement}, there may be scenarios where the elements of lists $R$ and $S$ are incomparable, making rule based blocking methods infeasible. To make a case for our method for such settings, we also evaluate \sysname\ against baselines approaches on a multilingual dataset \cite{multilingual}. Section \ref{sec:multilingual} provides more information on the dataset, as well as describes the corresponding experiments and results.

\paragraph{Evaluation Metrics}
We are interested in three questions to evaluate our matcher and blocker system:
\begin{itemize}
    \item Recall of the Blocker: What fraction of the duplicates $\dups$ are retrieved in $\cand$.
    \item Overall F1 score on unseen test pairs:
    How accurately can our system classify unseen pairs from a test set, $\mathcal{D}_{\text{test}}$, into duplicates and non duplicates.  The overall system predicts a record pair to be a duplicate only if the record pair is retrieved in $\cand$, and the matcher assigns a probability greater than $0.5$ of the pair being a duplicate. 
    \item Overall F1 score on all pairs: How accurately can our system find all duplicate pairs from the set of all possible pairs in the data? We compare the gold list of all duplicates in the data, to pairs that our system predicts to be duplicates. 
\end{itemize}
The test dataset,
$\mathcal{D}_{\text{test}}$, is the same dataset used to evaluate DeepMatcher. Hence, the test set evaluation metric gives us a way to compare our system with other approaches that may or may not be using active learning. However, the evaluation on all pairs is more aligned with the practical utility of any EM system. 

\subsection{Implementation Details}
\label{sec:implementation}
\subsubsection*{Compute}
We implemented all the systems, and experiments, in PyTorch $1.6$ \cite{pytorch}, and used transformers library by huggingface \cite{huggingface}. All experiments were conducted on a machine with $64$ 2.10GHz Intel Xeon Silver 4216 CPUs with $1007$GB RAM and a single NVIDIA Titan Xp $12$ GB GPU with CUDA $10.2$ running Ubuntu $18.04$. To retrieve nearest neighbours we use the Facebook AI Similarity Search (FAISS)~\cite{faiss} library. 

\subsubsection*{Model Architectures}
 We use the pre-trained RoBERTa model as our base transformer. The RoBERTa model builds on BERT, but with a careful selection of training sensitive hyperparameters like learning rate, and batch size. 
 The RoBERTa model was pre-trained on five English corpora of 160GB total size, 10 times that used for BERT. 
We use 6 layers out of the 12 layered uncased RoBERTa base model.
We use $12$ attention heads, with $768$ dimensional hidden vectors, and limit the number of input tokens to $512$. 
The paired classifier, on top of the base RoBERTa model, is the default classification head used in RoBERTa based models, consisting of two dropout layers with dropout probability $0.1$, a fully connected layer with a tanh activation, and a softmax classifier layer. Unless stated otherwise, \sysname\ uses a committee of size $\numC=3$, with each member using a masking probability $p=0.5$. 

\subsubsection*{Optimization}
We use the AdamW (Adam with Weight Decay) optimizer \cite{adamw}, with a learning rate of $3\mathrm{e}-5$ for the base transformer, and $1\mathrm{e}-3$ for the embedding and classifier layers. We use a linear learning rate schedule with no warm-up steps. The choice of optimizer parameters, and learning rate schedule, was based on previous works \cite{ditto, DBLP:conf/edbt/BrunnerS20}, and the standard choices for using RoBERTa models for classification tasks. We did not tune these hyper-parameters. The mini-batch size is $16$. The number of epochs is $20$ for the matcher and $200$ for the blocker.
\subsubsection*{Active Learning}
We conduct $10$ rounds of active learning, with a labeling budget of $B = 128$ samples per round. We start with an initial labeled seed set containing $|\train_p| = 64$ positive and $|\train_n| = 64$ negative pairs. These pairs were sampled at random from the benchmarked training splits of the datasets. All results are averaged over three such randomly constructed labeled seed sets. The default value of the candidate set size is $|\cand| = 3 \cdot |S|$ where $|S|$ denotes the size of the second list. The number of nearest neighbours retrieved is $k=3$. The size of List $S$, shown in Table \ref{tab:datasets}, is very small for the \abtbuy\ dataset, hence we use a candidate set size of $\cand = 20 \cdot |S|$, and $k=20$ for this dataset. We retrieve the nearest neighbours based on the $\ell_2$ distance. We do not warm start the model parameters between active learning rounds, i.e. after each round $M$ is re-initialized with the pre-trained weights of the TPLM. 

Unless stated otherwise, all systems use uncertainty sampling to select examples from the candidate set.  In all our experiments, we exclude the pairs in $\mathcal{D}_{\text{test}} \cap \cand$ from the process of selecting examples to query the labeler. 

\subsection{Methods Compared}
\label{sec:baselines}
We compare \sysname\ with four baseline approaches of blocking while using a TPLM-based matcher in an active learning loop:
\begin{itemize}
    \item {\bf \pairedFixed} uses a non adaptive blocking strategy, where the candidate set is created by conducting a similarity search on the embeddings obtained as is from the pre-trained TPLM, i.e. no task specific finetuning is employed
    \item {\bf \onlyPaired} uses the embeddings from the TPLM as it gets finetuned by the matcher in paired mode as described in Section \ref{sec:matcher}. However, the candidate set is created in a similar manner as \pairedFixed, i.e. a similarity search on the embeddings obtained from the TPLM in the single mode.
    \item {\bf \dittoAdapt} finetunes the TPLM and a SentenceBERT-like classifier on the labeled data $\train$ to obtain embeddings conducive for similarity search. 
    To keep comparisons uniform, even though the method is called SentenceBERT, we use the same RoBERTa transformer in all methods.  This method is also what is called the Advanced Blocking method in DITTO \cite{ditto} except that we learn it in an Active Learning setup much like \sysname. 
    \item {\bf \human} depends on hand-crafted rules to perform blocking. These exist only for the five benchmark datasets and not for the multilingual dataset. These five benchmarks already provide pairs 
    after pre-blocking with human-designed rules, so we did not create our own rules and instead define all pairs in these pre-blocked datasets as the candidate set for this method.
\end{itemize}
All baselines use a TPLM based matcher, similar to \sysname. 

We further compare \sysname\ with three well-established non-TPLM based methods. \citep{10.14778/2735471.2735474} conducted an exhaustive experimental study to compare various active learning methods for entity resolution on several real-world datasets and found that random forests with learner-aware QBC, described in Section \ref{sec:qbc}, perform remarkably well. We compare \sysname\ with a Random Forest learner implemented as an ensemble of $20$ decision trees using QBC via bootstrap~\citep{10.14778/2735471.2735474}.

JedAI~\citep{jedai2018,jedai2020} is another recent  open-source toolkit for Entity Resolution. JedAI offers highly scalable implementations of end-to-end schema-based and schema-agnostic pipelines including meta-blocking techniques. Schema-based workflows rely on similarity joins, whereas schema-agnostic workflows leverage all attribute values to extract overlapping blocks. We compare \sysname\ with the \textit{best configuration}~\citep{jedai2020} of both workflows, as found through Grid Search on each dataset using the gold list of duplicates $\dups$.






\begin{figure*}
\centering
\begin{tikzpicture}
\begin{groupplot}[group style={group size= 5 by 1,ylabels at=edge left, horizontal sep=0.75cm},width=0.24\textwidth,height=0.18\textwidth]
\nextgroupplot[label style={font=\normalsize},
xtick={500, 750, 1000, 1250},
tick label style={font=\normalsize},
legend style={at={($(0,0)+(1cm,1cm)$)},legend columns=4,fill=none,draw=black,anchor=center,align=center},
ylabel = {F1 (in \%)},
legend to name=fredf1,
title = \walmartamazon,
mark size=1pt]
\addplot [blue,mark=diamond*] coordinates {
(512,50.80246160965777) (640,61.21208108008682) (768,64.2811304834375) (896,55.195632393084615) (1024,62.72064868875871) (1152,58.64721864721864) (1280,59.400838544960536)};
\addplot[red!70!black,mark=*] coordinates {
(512, 70.5158599895442) (640, 73.4178294007353) (768, 75.3090626640946) (896, 75.5759702127835) (1024, 78.0619158232398) (1152, 75.81879113794) (1280, 76.7095718885864)};
\addplot[brown,mark=star] coordinates {
(512, 65.6101093730748) (640, 67.9351609564392) (768, 70.2757204308925) (896, 69.3577082913922) (1024, 71.1555986712526) (1152, 70.7121255968865) (1280, 68.0964603334064)};
\addplot[cyan,mark=square*] coordinates {
(512, 76.7367286690997) (640, 79.5076054937118) (768, 78.349450245698) (896, 82.0812385161684) (1024, 82.8528137433822) (1152, 81.5300594926732) (1280, 82.9665626311408)};
\addlegendentry{\dittoAdapt}
\addlegendentry{\pairedFixed};    
\addlegendentry{\onlyPaired};    
\addlegendentry{\sysname};          
\coordinate (c1) at (rel axis cs:0,1);

\nextgroupplot[label style={font=\normalsize},
xtick={500, 750, 1000, 1250},
tick label style={font=\normalsize},
title = \amazongoogle,
mark size=1pt]
\addplot [blue,mark=diamond*] coordinates {
(512,47.2750260300461) (640,48.12026325716114) (768,51.89662455333403) (896,52.27652626573936) (1024,51.93380278094261) (1152,46.15133423641564) (1280,51.44022125267518)};
\addplot [red!70!black,mark=*] coordinates {
(512, 59.0924469806168) (640, 58.6446035489897) (768, 60.1904801698771) (896, 61.5472356778953) (1024, 61.3754614006582) (1152, 61.7926717829861) (1280, 61.2243894856519)};
\addplot [brown,mark=star] coordinates {
(512, 59.1677564746349) (640, 59.5306122365388) (768, 62.1693060095282) (896, 63.6734492530906) (1024, 62.6851769608246) (1152, 64.0399768376347) (1280, 61.5743727459964)};
\addplot [cyan,mark=square*] coordinates {
(512, 63.3634220211925) (640, 62.5057243555198) (768, 65.2954025379654) (896, 67.7894787114645) (1024, 68.9767580784085) (1152, 69.3575292551642) (1280, 69.2116052903128)};

\nextgroupplot[label style={font=\normalsize},
xtick={500, 750, 1000, 1250},
tick label style={font=\normalsize},
title = \dblpacm,
mark size=1pt]
\addplot [blue,mark=diamond*] coordinates {
(512,94.67547454387476) (640,96.47881007564264) (768,96.22651476286129) (896,97.29335850895251) (1024,93.3428910325242) (1152,96.59255464034217) (1280,96.92405777041674)};
\addplot [red!70!black,mark=*] coordinates {
(512, 96.1849302131182) (640, 96.1824471428482) (768, 96.1820274031256) (896, 96.5516930916817) (1024, 96.4024865423995) (1152, 96.4024865423995) (1280, 96.3734535858484)};
\addplot [brown,mark=star] coordinates {
(512, 98.3292280283405) (640, 98.5614614795405) (768, 98.2961824368455) (896, 98.3301940150583) (1024, 98.6738533150653) (1152, 98.4426856847673) (1280, 98.5980083752727)};
\addplot [cyan,mark=square*] coordinates {
(512, 98.302861625742) (640, 98.4870817013674) (768, 98.5244040862655) (896, 98.7159831845157) (1024, 98.600838897449) (1152, 98.7505566918048) (1280, 98.7897673638261)};
\nextgroupplot[label style={font=\normalsize},
xtick={500, 750, 1000, 1250},
tick label style={font=\normalsize},
title = \dblpscholar,
mark size=1pt]
\addplot [blue,mark=diamond*] coordinates {
(512,83.06136446211609)
(640,76.09257511772847)
(768,86.83540208421556)
(896,87.05448392751197)
(1024,85.12615866340218)
(1152,84.2120775698639)
(1280,83.4612444642799)};
\addplot [red!70!black,mark=*] coordinates {
(512,83.62722863269526)
(640,83.41261534659346)
(768,83.58283848779475)
(896,83.79317622466388)
(1024,83.79095837528506)
(1152,83.88345898718303)
(1280,83.95836796673132)};
\addplot [brown,mark=star] coordinates {
(512,88.85686624828836)
(640,90.22334966176459)
(768,90.33292921075126)
(896,90.9175979111918)
(1024,89.50600253944853)
(1152,90.61530935802014)
(1280,89.9800518174708)};
\addplot [cyan,mark=square*] coordinates {
(512,92.84526676698327)
(640,93.4687077908332)
(768,93.6993628251506)
(896,94.08584779482133)
(1024,94.1637109816327)
(1152,94.72660513428602)
(1280,94.82725838488707)};
\nextgroupplot[label style={font=\normalsize},
xtick={500, 750, 1000, 1250},
tick label style={font=\normalsize},
title = \abtbuy,
mark size=1pt]
\addplot [blue,mark=diamond*] coordinates {
(512,19.472907442546912)
(640,28.96973487069246)
(768,38.44560875579119)
(896,44.39409937764836)
(1024,39.7389009457975)
(1152,48.144851673656675)
(1280,29.06547282183954)};
\addplot [red!70!black,mark=*] coordinates {
(512,45.11406210999579)
(640,44.23834493037826)
(768,46.37276344567709)
(896,45.28469318613323)
(1024,45.300665392510794)
(1152,46.280858968132634)
(1280,45.65059183660332)};
\addplot [brown,mark=star] coordinates {
(512,37.9438602254254)
(640,34.589080196983154)
(768,33.60739246490209)
(896,29.58555736835391)
(1024,33.44270148302513)
(1152,35.503880475098185)
(1280,34.45339285534743)};
\addplot [cyan,mark=square*] coordinates {
(512,81.11372543843149)
(640,81.73424847985915)
(768,84.05724414258033)
(896,86.4311952821263)
(1024,86.81148370349136)
(1152,88.81004775468239)
(1280,86.44004203938742)};

\coordinate (c2) at (rel axis cs:1,1);

\end{groupplot}
\coordinate (c3) at ($(c1)!.5!(c2)$);
\node[below] at (c3 |- current bounding box.south)
{\pgfplotslegendfromname{fredf1}};

\end{tikzpicture}
\caption{Comparison of \sysname\ with baseline approaches with respect to F1 on a fixed test-set against increasing number of instances selected by active learning. In all cases, \sysname\ provides significant gains over existing methods.}
\label{fig:maintestf1}
\end{figure*}
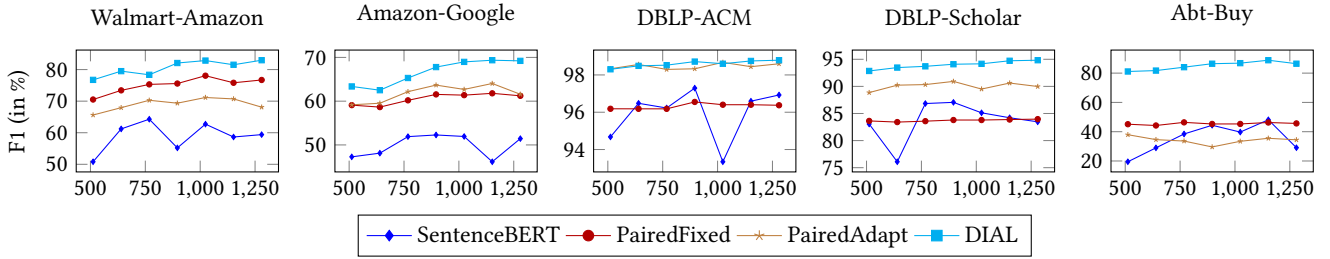

\setlength\tabcolsep{3.0pt}
\begin{table*}
    \centering
           \caption{Comparison of \sysname\ with baseline approaches with respect to Precision, Recall, and F1 evaluated on all pairs at the end of the AL loop. \sysname\ achieves high recall and consequently high F1 scores. The RT column denotes time in seconds to find All duplicate pairs and includes both blocking and matching time.}
    \begin{tabular}{l|c|c|c|c|c|c|c|c|c|c|c|c|c|c|c|c|c|c|c|c}
        \toprule
        \multirow{2}{*}{Method} & \multicolumn{4}{c}{\walmartamazon} & \multicolumn{4}{|c}{\amazongoogle} & \multicolumn{4}{|c|}{\dblpacm} & \multicolumn{4}{|c|}{\dblpscholar}  & \multicolumn{4}{|c}{\abtbuy}\\
        \cline{2-21}
        & P & R & F1 & RT & P & R & F1 & RT & P & R & F1 & RT & P & R & F1  & RT & P & R & F1 & RT \\
        \midrule
        \multicolumn{21}{c}{Non TPLM based}\\
        \midrule
        Random Forest & $96.5$ & $63.0$ & $76.2$ & $1.1$ & $84.7$ & $54.6$ & $66.3$ & $1.1$ & $99.0$ & $99.1$ & $99.0$ & $1.3$ & $97.2$ & $96.3$ & $96.7$ & $2.7$ & $83.9$ & $52.4$ & $64.4$ & $0.9$\\
        JedAI:Schema-based & $82.9$ & $55.2$ & $66.3$ & $0.5$ & $66.3$ & $42.3$ & $51.7$ & $0.5$ & $97.8$ & $93.2$ & $95.4$ & $0.6$ & $95.3$ & $77.5$ & $85.5$ & $14$ & $88.4$ & $43.8$ & $58.5$ & $0.4$\\
        JedAI:Schema-agnostic & $59.0$ & $75.3$ & $66.2$ & $5.3$ & $57.6$ & $64.1$ & $60.7$ & $4.5$ & $99.3$ & $\mathbf{99.2}$ & $\mathbf{99.3}$ & $1.3$ & $94.6$ & $94.9$ & $94.7$ & $30$ & $94.9$ & $85.6$ & $90.0$ & $1.1$\\
        \midrule
        \multicolumn{21}{c}{TPLM based}\\
        \midrule
        \dittoAdapt & $87.1$ & $43.9$ & $58.0$ & $87.6$ & $73.2$ & $38.5$ & $50.4$ & $7.9$ & $99.3$ & $94.3$& $96.7$ & $15.5$ & $97.0$ & $74.4$ & $84.2$ & $255$ & $87.6$ & $20.3$ & $32.6$ & $42$\\
        \pairedFixed & $96.6$ & $71.2$ & $82.0$ & $87.6$ & $94.9$ & $52.1$ & $67.2$ & $7.9$ & $99.6$ & $93.6$ & $96.5$ & $15.5$ & $98.5$ & $74.2$ & $84.6$ & $255$ & $97.9$ & $33.0$ & $49.3$ & $42$\\
        \onlyPaired & $96.3$ & $61.2$ & $74.4$ & $87.6$ & $91.6$ & $58.3$ & $71.1$ & $7.9$ & $99.7$ & $98.0$ & $98.8$ & $15.5$ & $98.2$ & $85.8$ & $91.6$ & $255$ & $97.6$ & $23.4$ & $37.7$ & $42$\\
        \human & $93.7$ & $77.3$ & $84.7$ & $9.2$ & $85.4$ & $75.2$ & $79.9$ & $5.6$ & $99.4$ & $\mathbf{99.2}$ & $\mathbf{99.3}$ & $15.1$ & $96.3$ & $\mathbf{98.0}$ & $\mathbf{97.1}$ & $26$ & $96.3$ & $87.2$ & $91.6$ & $15$\\
\sysname & $94.9$ & $\mathbf{85.2}$ & $\mathbf{89.8}$  & $88.3$ & $87.4$ & $\mathbf{77.4}$ & $\mathbf{82.1}$ & $8.0$ & $99.6$ & $98.6$ & $99.1$ & $15.6$ & $97.5$ & $96.1$ & $96.8$ & $257$ & $97.8$ & $\mathbf{87.4}$ & $\mathbf{92.3}$ & $42$\\
        \bottomrule
    \end{tabular}

    \label{tab:allpairsal}
\end{table*}
\setlength\tabcolsep{5.0pt}

\begin{figure*}
\centering
\begin{tikzpicture}
\begin{groupplot}[group style={group size= 5 by 1,ylabels at=edge left, horizontal sep=0.75cm},width=0.24\textwidth,height=0.18\textwidth]
\nextgroupplot[label style={font=\normalsize},
xtick={500, 750, 1000, 1250},
tick label style={font=\normalsize},
legend style={at={($(0,0)+(1cm,1cm)$)},legend columns=5,fill=none,draw=black,anchor=center,align=center},
ylabel = {Recall (in \%)},
legend to name=fredrecall,
title = \walmartamazon,
mark size=1pt]
\addplot [blue,mark=diamond*] coordinates {
(512,43.35644136337377)
(640,56.672443674176776)
(768,67.67764298093587)
(896,47.631426920855006)
(1024,58.69439630271519)
(1152,51.27094165222415)
(1280,50.14442518775275)};
\addplot[red!70!black,mark=*] coordinates {
(512,74.95667244367418)
(640,74.95667244367418)
(768,74.95667244367418)
(896,74.95667244367418)
(1024,74.95667244367418)
(1152,74.95667244367418)
(1280,74.95667244367418)};
\addplot[brown,mark=star] coordinates {
(512,64.44251877527441)
(640,64.00924321201616)
(768,66.8110918544194)
(896,65.9734257654535)
(1024,66.66666666666666)
(1152,66.72443674176776)
(1280,64.84690930098209)};
\addplot [gray,mark=triangle*] coordinates {
(512,83.36221837088388)
(640,83.36221837088388)
(768,83.36221837088388)
(896,83.36221837088388)
(1024,83.36221837088388)
(1152,83.36221837088388)
(1280,83.36221837088388)};
\addplot[cyan,mark=square*] coordinates {
(512,83.50664355863663)
(640,86.22183708838821)
(768,87.92605430387059)
(896,89.45696129404968)
(1024,90.90121317157713)
(1152,91.82553437319469)
(1280,92.2010398613518)};
\addlegendentry{\dittoAdapt}
\addlegendentry{\pairedFixed};    
\addlegendentry{\onlyPaired};    
\addlegendentry{\human};      
\addlegendentry{\sysname};      
\coordinate (c1) at (rel axis cs:0,1);

\nextgroupplot[label style={font=\normalsize},
xtick={500, 750, 1000, 1250},
tick label style={font=\normalsize},
title = \amazongoogle,
mark size=1pt]
\addplot [blue,mark=diamond*] coordinates {
(512,38.64102564102564)
(640,42.89743589743589)
(768,50.46153846153847)
(896,48.41025641025641)
(1024,48.769230769230774)
(1152,42.07692307692308)
(1280,46.8974358974359)};
\addplot [red!70!black,mark=*] coordinates {
(512,56.46153846153846)
(640,56.46153846153846)
(768,56.46153846153846)
(896,56.46153846153846)
(1024,56.46153846153846)
(1152,56.46153846153846)
(1280,56.46153846153846)};
\addplot [brown,mark=star] coordinates {
(512,62.487179487179475)
(640,64.71794871794872)
(768,63.4102564102564)
(896,66.17948717948718)
(1024,65.1025641025641)
(1152,62.846153846153854)
(1280,63.871794871794876)};
\addplot [gray,mark=triangle*] coordinates {
(512,89.76923076923077)
(640,89.76923076923077)
(768,89.76923076923077)
(896,89.76923076923077)
(1024,89.76923076923077)
(1152,89.76923076923077)
(1280,89.76923076923077)};
\addplot [cyan,mark=square*] coordinates {
(512,80.12820512820512)
(640,82.25641025641025)
(768,83.84615384615385)
(896,84.58974358974359)
(1024,86.46153846153845)
(1152,87.07692307692308)
(1280,88.35897435897436)};

\nextgroupplot[label style={font=\normalsize},
xtick={500, 750, 1000, 1250},
tick label style={font=\normalsize},
title = \dblpacm,
mark size=1pt]
\addplot [blue,mark=diamond*] coordinates {
(512,93.54016786570742)
(640,96.67266187050359)
(768,94.88908872901679)
(896,96.71762589928056)
(1024,89.52338129496404)
(1152,95.51858513189448)
(1280,94.9040767386091)};
\addplot [red!70!black,mark=*] coordinates {
(512,93.70503597122303)
(640,93.70503597122303)
(768,93.70503597122303)
(896,93.70503597122303)
(1024,93.70503597122303)
(1152,93.70503597122303)
(1280,93.70503597122303)};
\addplot [brown,mark=star] coordinates {
(512,97.72182254196643)
(640,98.41127098321344)
(768,98.18645083932854)
(896,98.44124700239809)
(1024,98.56115107913669)
(1152,98.11151079136691)
(1280,98.3962829736211)};
\addplot [gray,mark=triangle*] coordinates {
(512,99.82014388489209)
(640,99.82014388489209)
(768,99.82014388489209)
(896,99.82014388489209)
(1024,99.82014388489209)
(1152,99.82014388489209)
(1280,99.82014388489209)};
\addplot [cyan,mark=square*] coordinates {
(512,98.66606714628297)
(640,98.6810551558753)
(768,98.81594724220624)
(896,98.95083932853717)
(1024,98.96582733812949)
(1152,98.98081534772182)
(1280,98.98081534772182)};

\nextgroupplot[label style={font=\normalsize},
xtick={500, 750, 1000, 1250},
tick label style={font=\normalsize},
title = \dblpscholar,
mark size=1pt]
\addplot [blue,mark=diamond*] coordinates {
(512,74.27841157035097)
(640,64.21669471978056)
(768,80.74932984227917)
(896,82.94370675144941)
(1024,78.28065581946262)
(1152,75.57508883486067)
(1280,76.11744903684308)};
\addplot [red!70!black,mark=*] coordinates {
(512,74.60258088647839)
(640,74.60258088647839)
(768,74.60258088647839)
(896,74.60258088647839)
(1024,74.60258088647839)
(1152,74.60258088647839)
(1280,74.60258088647839)};
\addplot [brown,mark=star] coordinates {
(512,84.76404214201109)
(640,86.08565550776136)
(768,86.57814350726264)
(896,87.12050370924507)
(1024,85.04457328096753)
(1152,86.87114269683934)
(1280,86.48463312761051)};
\addplot [gray,mark=triangle*] coordinates {
(512,100)
(640,100)
(768,100)
(896,100)
(1024,100)
(1152,100)
(1280,100)};
\addplot [cyan,mark=square*] coordinates {
(512,93.58518795586309)
(640,94.78212081541052)
(768,95.46786359952621)
(896,95.85437316875506)
(1024,96.5151798516302)
(1152,97.13234835733432)
(1280,97.30066704070818)};
\nextgroupplot[label style={font=\normalsize},
xtick={500, 750, 1000, 1250},
tick label style={font=\normalsize},
title = \abtbuy,
mark size=1pt]
\addplot [blue,mark=diamond*] coordinates {
(512,7.444545730780917)
(640,8.052263749620177)
(768,17.988453357642054)
(896,17.046490428441206)
(1024,21.05742935278031)
(1152,13.704041324825281)
(1280,13.643269522941356)};
\addplot [red!70!black,mark=*] coordinates {
(512,25.52415679124886)
(640,25.52415679124886)
(768,25.52415679124886)
(896,25.52415679124886)
(1024,25.52415679124886)
(1152,25.52415679124886)
(1280,25.52415679124886)};
\addplot [brown,mark=star] coordinates {
(512,24.551807961106046)
(640,23.488301428137344)
(768,22.759039805530236)
(896,21.756305074445457)
(1024,20.297781829231234)
(1152,26.101488909146152)
(1280,21.422060164083863)};
\addplot [gray,mark=triangle*] coordinates {
(512,93.71011850501367)
(640,93.71011850501367)
(768,93.71011850501367)
(896,93.71011850501367)
(1024,93.71011850501367)
(1152,93.71011850501367)
(1280,93.71011850501367)};
\addplot [cyan,mark=square*] coordinates {
(512,77.14980249164388)
(640,80.21877848678213)
(768,82.43694925554543)
(896,84.19933151017928)
(1024,85.26283804314798)
(1152,85.3236098450319)
(1280,86.53904588271043)};

\coordinate (c2) at (rel axis cs:1,1);

\end{groupplot}
\coordinate (c3) at ($(c1)!.5!(c2)$);
\node[below] at (c3 |- current bounding box.south)
{\pgfplotslegendfromname{fredrecall}};

\end{tikzpicture}
\caption{Recall on $\cand$ against increasing number of instances selected by active learning. In all cases, \sysname\ provides significant gains over baseline methods and is able to achieve recall at par with hand crafted rule based blocking.}
\label{fig:recall}
\end{figure*}
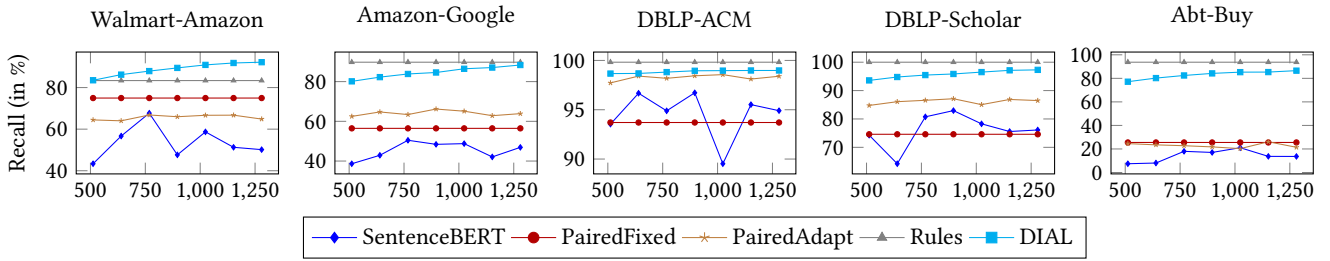
\subsection{Overall Results}
\label{sec:expBlocker}
Figure \ref{fig:maintestf1} plots the progressive F1 scores obtained by overall system of baseline methods and \sysname\, on the unseen test dataset as described in Section \ref{sec:datasets}. The x-axis denotes the increasing number of example pairs in $\train$ as active learning progresses.  In Table~\ref{tab:allpairsal} we show the efficacy of each system at the end of AL in retrieving all duplicate pairs.  Here we also show precision, recall and running time.
We find that \sysname\ provides significant gains in F1 over baselines methods both at each stage of AL and at the end of AL. 



Table~\ref{tab:allpairsal} shows that \sysname~produces the best F1 scores on all the product datasets while performing close to the best on the citation datasets. With respect to recall, we note that \sysname's recall is often close and in some cases, perhaps surprisingly, \emph{exceeds} \human'. The intent behind \human~was to perform blocking which in turn, calls for recall. So it is quite surprising that on datasets such as \walmartamazon~and \abtbuy, without any external knowledge about the domain and with only limited number of judiciously chosen labels, \sysname~produces even better recall than hand-tuned blocking functions. 
Figure~\ref{fig:recall} provides a more detailed view of this phenomenon by showing the recall of the candidate set \cand\ at each stage of AL.  Here we see that the recall offered by \sysname's blocker is significantly higher than other methods.  Note the recall of \pairedFixed\ and \human\ does not change since the candidate set remains fixed.  In most cases \onlyPaired's F1 is better than \pairedFixed\ indicating that fine-tuning the transformer parameters with the task specific training data $\train$ is helpful.  The recall of \dittoAdapt\ is worse than \onlyPaired\ perhaps because the SentenceBert network architecture, choice of training data, and training objective are not effective in co-embedding duplicates. The finding on the poor performance of \dittoAdapt\  is significant because DITTO~\cite{ditto}, a recent state-of-the-art ER system proposed to use SentenceBERT as its advanced blocking strategy on their large internal dataset.

In terms of running time, we observe that all deep-learning (TPLM-based) methods are between one and two orders of magnitude slower than pre-deep learning methods in the first three rows.  However, given the substantial gains in accuracy that TPLM-based methods provide, an end-user may be willing to invest in the extra running time. 

\begin{table}
    \centering
          \caption{Precision, Recall, and F1 evaluated on all pairs on the Multilingual dataset at the end of the $10$ AL rounds. \sysname\ achieves higher almost 7.3 points higher F1 compared to existing practice of solving this task.}
    \begin{tabular}{l|c|c|c}
        \toprule
        Method &  P & R & F1\\
        \midrule
        \pairedFixed & $81.2$ & $56.8$ & $66.9$ \\
        \onlyPaired  & $94.8$ & $31.6$ & $47.4$\\
        \sysname & $92.2$ & $\mathbf{62.3}$ & $\mathbf{74.3}$\\
        \bottomrule
    \end{tabular}
\label{tab:allpairsmultilingual}
\end{table}

\begin{figure}
\centering
\begin{tikzpicture}
\begin{groupplot}[group style={group size= 1 by 1,ylabels at=edge left},width=0.33\textwidth,height=0.24\textwidth]
\nextgroupplot[label style={font=\normalsize},
xtick={500, 750, 1000, 1250},
tick label style={font=\normalsize},
legend style={at={($(0,0)+(1cm,1cm)$)},legend columns=6,fill=none,draw=black,anchor=center,align=center},
ylabel = {F1 (in \%)},
legend to name=fredmulti,
title = \multilingual,
mark size=1pt]
\addplot [red!70!black,mark=*] coordinates {
(512,69.54) (640,71.71) (768,71.96) (896,73.98) (1024,72.45) (1152,72.12) (1280,74.61)};
\addplot [brown,mark=star] coordinates {
(512,51.41158989598811) (640,51.19047619047619) (768,59.55056179775281) (896,66.22073578595317) (1024,62.448418156808806) (1152,47.44469870327994) (1280,52.616064848931465)};
\addplot [cyan,mark=square*] coordinates {
(512,75.16) (640,72.37) (768,80.88) (896,79.59) (1024,79.95) (1152,81.66) (1280,80.24)};
\addlegendentry{\pairedFixed};
\addlegendentry{\onlyPaired};
\addlegendentry{\sysname};
\coordinate (c2) at (rel axis cs:1,1);

\end{groupplot}
\coordinate (c3) at ($(c1)!.5!(c2)$);
\node[below] at (c3 |- current bounding box.south)
{\pgfplotslegendfromname{fredmulti}};

\end{tikzpicture}
\caption{Comparison of \sysname\ with baselines on progressive F1 scores on a fixed test-set. \sysname\ consistently outperforms baseline methods.}
\label{fig:multilingual}
\end{figure}
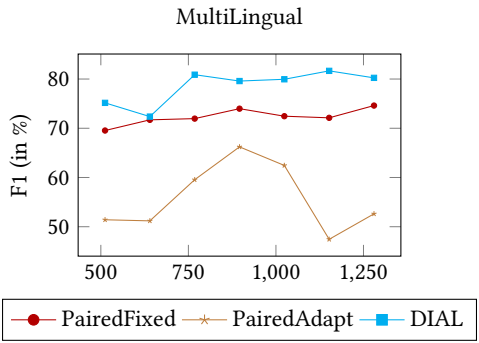

\subsection{Multilingual Dataset}
\label{sec:multilingual}
The multilingual dataset that we use is from \cite{multilingual}. The dataset, originally proposed for machine translation of structured data, consists of accurately-aligned parallel XML files in multiple languages. For our experiments, we use the English-Deutsch subset. Concretely, in our setup, each element of list $R$ is a string in English which can contain HTML/XML tags, and similarly each element of list $S$ is a string in German which can contain HTML/XML tags. As a result of the parallel alignment in data, we have $|\dups|=|R|=|S|$. 

For the multilingual dataset, we use 6 layers out of the 12 layered uncased multilingual BERT base model. This model was pre-trained on 104 languages from the Wikipedia dataset using Masked Language Modelling and Next Sentence Prediction \cite{devlin-etal-2019-bert}. Apart from changing the base transformer all other implementation details, including the architectures of the classifiers remain the same as that for the earlier five benchmark datasets.

We now describe the construction of the labeled seed set. We use a pre-trained $12$ layered uncased multilingual BERT base model and create an index on the embedding of each $r \in R$. Then, we query $k=3$ nearest neighbours in this index for the embedding of each $s \in S$. Using the gold list of duplicates $\dups$, we divide these retrieved pairs into duplicates and non-duplicates. A random sample of $64$ duplicate, and $64$ non-duplicate pairs, from these sets respectively, is then chosen to create the labeled seed set. The test set is constructed in a similar manner, except the index is created, and probed, on the elements of the dev split of the dataset. 
The multilingual BERT model, as mentioned above, is pre-trained on $104$ different languages and hence learns an extremely strong prior. Moreover, the dataset that we use consists mostly of natural language text, as opposed to the deepmatcher datasets involving product or bibliographical data. These two key  differences from the previous setup influence the decision to fine-tune the TPLM, i.e. we find that freezing the TPLM parameters leads to slightly better F1 scores. 

Progressive F1 scores calculated on the test data can be found in Figure \ref{fig:multilingual}. Table \ref{tab:allpairsmultilingual} compares \sysname\ against the \pairedFixed\ and \onlyPaired\ baselines on All-Pairs F1 scores calculated after $10$ active learning rounds. We notice that on both evaluation measures, \sysname\ outperforms baselines significantly. Compared to indexing the transformer embeddings as-is, \sysname\ achieves more than 7 percent points increase in F1!

\begin{table}
    \centering
           \caption{Comparing labeled negatives with random negatives to train the committee embeddings in \sysname\ after $10$ rounds of AL.  Note the 12-25 points jump in recall with Random negatives on product datasets.}
    \begin{tabular}{c|c|c|c|c|c}
        \toprule
        Negatives & W-A & A-G & D-A & D-S & A-B\\
        \midrule
        \multicolumn{6}{c}{Recall of $\cand$}\\
        \midrule
        Labeled & $80.94$ & $76.54$ & $\mathbf{99.02}$ & $93.47$ & $66.45$\\
        Random & $\mathbf{92.20}$ & $\mathbf{88.36}$ & $\mathbf{98.98}$ & $\mathbf{97.30}$ & $\mathbf{92.50}$\\
        \midrule
        \multicolumn{6}{c}{Test Evaluation}\\
        \midrule
        Labeled & $75.47$ & $67.93$ & $\mathbf{98.75}$ & $93.32$ & $69.74$\\
        Random & $\mathbf{82.97}$ & $\mathbf{69.21}$ & $\mathbf{98.79}$ & $\mathbf{94.83}$ & $\mathbf{88.81}$\\
        \midrule
        \multicolumn{6}{c}{All Pairs Evaluation}\\
        \midrule
        Labeled & $85.36$ & $78.78$ & $\mathbf{99.14}$ & $95.49$ & $78.12$\\
        Random & $\mathbf{89.80}$ & $\mathbf{82.07}$ & $\mathbf{99.13}$ & $\mathbf{96.81}$ & $\mathbf{92.31}$\\
        \bottomrule
    \end{tabular}
    \label{tab:negatives}
\end{table}

\subsection{Ablation Study}
We next present a detailed ablation study to evaluate the impact of the many design decisions we made in the design of \sysname.
\subsubsection{Choice of Training Data}
To validate the intuition presented in Section \ref{sec:blockertrainingdata},  we compare \sysname, where the blocker is trained to drive apart embeddings of ``easy" non-duplicates, to where the blocker is trained to separate the ``difficult"  labeled negatives ($T-T_p$) chosen during \sysname's AL loop. Table \ref{tab:negatives} evaluates, on all three metrics, the two systems.
We observe that Random Negatives achieves higher recall on $\cand$ providing absolute gains of $12-25\%$ over Labeled negatives on the product datasets! This subsequently results in much better F1 scores on both evaluation measures, compared to Labeled Negatives. We note here that Random Negatives while significantly improving recall of blocker, can be detrimental to precision if used to train the Matcher. On product datasets, a matcher trained with random negatives suffers a loss of $30-60\%$ in precision compared to labeled negatives.

\subsubsection{Choice of Training Objective}
Once we have established that random negatives are more effective than labeled negatives, we evaluate the  objective function to train the blocker for maximizing recall. We compare our Contrastive objective, defined in Equation~\ref{eq:contrast}, with two other objectives: 

\noindent {\bf Classification} objective as used in SentenceBert to separate duplicates from non-duplicates using cross entropy (Eq~\ref{eq:matcher})

\noindent
{\bf Triplet} Used in \cite{tracz-etal-2020-bert} for product matching with TPLM, a triplet loss is computed on examples that are positive and negative with respect to an anchor. This loss penalizes the model if the anchor is farther away from the positive than the negative example. The TripletObjective is expressed as
\begin{align*}
  \text{TripletObjective} &= \max(d(r_p, s_p) - d(r_p, s_r) + \text{margin}, 0))\\
  &+ \max(d(s_p, r_p) - d(s_p, r_r) + \text{margin}, 0))  
\end{align*}
We use the euclidean distance metric $d$, and set the margin to be $1$. However, unlike \cite{tracz-etal-2020-bert}, we do not perform hard negative mining. 

\begin{table}
    \centering
           \caption{Evaluation of \sysname\ with different objectives to train the committee embeddings after $10$ rounds of Active Learning. Contrastive objective consistently outperforms Classification and Triplet objectives.}
    \begin{tabular}{l|c|c|c|c|c}
        \toprule
        Objective & W-A & A-G & D-A & D-S & A-B\\
        \midrule
        \multicolumn{6}{c}{Test Evaluation}\\
        \midrule
        Classification & $79.63$ & $67.40$ & $98.75$ & $93.28$  & $70.90$\\
        Triplet & $80.94$ & $68.71$  & $\mathbf{98.79}$ & $94.38$ & $87.21$\\
        Contrastive & $\mathbf{82.97}$ & $\mathbf{69.21}$ & $\mathbf{98.79}$ & $\mathbf{94.83}$ & $\mathbf{88.81}$\\
        \midrule
        \multicolumn{6}{c}{All Pairs Evaluation}\\
        \midrule
        Classification & $84.88$ & $79.17$ & $99.05$ & $95.15$ & $76.03$\\
        Triplet & $87.72$ & $81.04$ & $99.06$ & $96.48$ & $91.95$\\
        Contrastive & $\mathbf{89.80}$ & $\mathbf{82.07}$ & $\mathbf{99.13}$ & $\mathbf{96.81}$ & $\mathbf{92.31}$\\
        \bottomrule
    \end{tabular}
\label{tab:trainingobjective}
\end{table}

Table \ref{tab:trainingobjective} reports the F1 scores on Test and All pairs evaluations at the end of the active learning loop, for the three different training objectives used to train the blocker. We see that the Contrastive  consistently outperforms Classification and Triplet objectives. The similarity between instance embeddings of the positive (and negative) pairs is maximized (and minimized) explicitly in contrastive and triplet objectives, whereas this is implicit in classification. The contrastive objective is able to leverage multiple random negatives as opposed to triplet which only uses $2$, one for each instance as an anchor.

\subsubsection{Choice of Candidate Size}
\begin{table}
    \centering
            \caption{Evaluation of \sysname\ with increasing candidate set size after $10$ rounds of Active Learning.}
    \begin{tabular}{r|c|c|c|c|c}
        \toprule
        |\cand| & W-A & A-G & D-A & D-S & A-B\\
        \midrule
        \multicolumn{6}{c}{Recall}\\
        \midrule
        Small & $55.78$ & $79.31$ & $98.98$ & $92.55$ & $71.92$\\
        Medium & $92.20$ & $88.36$ & $98.98$ & $97.30$ & $86.54$\\
        Large & $94.60$ & $89.90$ & $99.09$ & $97.85$ & $92.50$\\
        \midrule
        \multicolumn{6}{c}{All Pairs Evaluation}\\
        \midrule
        Small & $70.19$ & $80.09$ & $99.08$ & $95.01$ & $82.68$\\
        Medium & $89.80$ & $82.07$ & $99.13$ & $96.81$ & $90.49$\\
        Large & $90.80$ & $81.41$ & $99.19$ & $97.00$ & $92.31$\\
        \bottomrule
    \end{tabular}

\label{tab:candidatesetsize}
\end{table}

The size of Candidate set $|\cand|$, is an important factor that influences the overall recall of the system. A small candidate set can lead to low recall, and a large candidate set can inadvertently lead to low precision. Table \ref{tab:candidatesetsize} compares \sysname\ with different candidate set sizes. Small corresponds to $|\cand| = 3\cdot|\dups|$. Medium and Large correspond to $|\cand| = 10\cdot|S|$ and $20\cdot|S|$ for \abtbuy\, and $|\cand| = 3\cdot|S|$ and $5\cdot|S|$ respectively for all other datasets. On average,  All Pairs Evaluation is maximized for Large $|\cand|$. 


\subsubsection{Impact of Committee size in our blocker}

\begin{table}
    \centering
            \caption{Evaluation of \sysname\ with increasing committee size ($\numC$) after $10$ rounds of Active Learning.}
    \begin{tabular}{c|c|c|c|c|c}
        \toprule
        $\numC$ & W-A & A-G & D-A & D-S & A-B\\
        \midrule
        \multicolumn{6}{c}{Test Evaluation}\\
        \midrule
        $1$ & $83.16$ & $68.62$ & $98.52$ & $94.38$ & $88.56$\\
        $3$ & $82.97$ & $69.21$ & $\mathbf{98.79}$ & $\mathbf{94.83}$ & $\mathbf{88.81}$\\
        $5$ & $\mathbf{83.51}$ & $\mathbf{70.85}$ & $98.71$ & $94.76$ & $88.31$\\
        \midrule
        \multicolumn{6}{c}{All Pairs Evaluation}\\
        \midrule
         $1$ & $89.85$ & $80.82$ & $\mathbf{99.20}$ & $96.21$ & $92.22$\\
         $3$ & $89.80$ & $82.07$ & $99.13$ & $\mathbf{96.81}$ & $92.31$\\
         $5$ & $\mathbf{90.19}$ & $\mathbf{82.14}$ & $99.10$ & $96.66$ & $\mathbf{92.79}$\\
        \bottomrule
    \end{tabular}
    \label{tab:committeesize}
\end{table}
Table \ref{tab:committeesize} evaluates \sysname\ with different committee sizes $\numC$. As we motivate in Section \ref{sec:committee}, the committee is introduced to improve recall with the intuition that as opposed to one embedding it is less likely that a duplicate pair is missed by a committee of different embeddings. We find that on average, having multiple members improves performance as compared to a single member
An immediate question that then arises is, what is the cost of introducing an additional member in the committee? In Section \ref{sec:runningtime} we provide a running time analysis varying the committee size. We show that \sysname\ is optimized to efficiently handle large committee sizes.

\subsection{Selection Strategies}
\label{sec:selection}
Unless stated otherwise, we have used Uncertainty Sampling as the example selection strategy for active learning. However, \sysname\ is agnostic to the choice of selection strategy. In this section, we compare different example selection strategies with \sysname. We implement the following methods
\begin{itemize}
    \item \textbf{Random}: The naive baseline of choosing samples at random from the candidate set
    \item \textbf{Greedy}: Selecting the most similar pairs from the candidate set. We use the negative $\ell_2$ distance as a similarity metric
    \item \textbf{Partition}: As explained in Section \ref{sec:exampleselection}, High Confidence Sampling with Partition is not strictly an Active Learning selection strategy since it assumes labels not provided by a human labeler. Hence, to use a similar method as \cite{DBLP:conf/acl/KasaiQGLP19} in our setup, we implement two selection strategies. Partition-$2$ queries the user to label $p_{lc}$ and $n_{lc}$, and Partition-$4$ queries the user to label $p_{hc}, p_{lc}, n_{hc}, n_{lc}$.  
    \item \textbf{Query By Committee}: Select pairs from the candidate set which achieve the highest disagreement in a committee of classifiers. If member $k$ of a committee of size $\numC$ assigns a probability $\Pr_k(y=1|(r, s))$ to a pair $(r, s)$ of being a duplicate pair, then the disagreement is measured as
    $$H\left(\frac{1}{\numC}{\sum\limits_{k=1}^{\numC} \Pr_k(y=1| (r, s))}\right)$$ where $H(x)$ is given by \myeqnref{eq:uncertain}. 
    Note that, this is a ``soft" measure of disagreement, as opposed to the the hard disagreement defined in Section \ref{sec:exampleselection}.
    \item \textbf{BADGE}: Described in Section \ref{sec:exampleselection}. For a record pair $(r, s) \in \cand$, the input $x$ is the joint encoding of $r$ and $s$. The most likely label $\hat{y}$ is calculated based on the class probabilities output by $F_W(E(x))$. The loss used to calculate the gradient embedding is the standard cross entropy loss.
\end{itemize}

\begin{figure*}
\centering
\begin{tikzpicture}
\begin{groupplot}[group style={group size= 5 by 1,ylabels at=edge left, horizontal sep=0.75cm},width=0.24\textwidth,height=0.18\textwidth]
\nextgroupplot[label style={font=\normalsize},
xtick={500, 750, 1000, 1250},
tick label style={font=\normalsize},
legend style={at={($(0,0)+(1cm,1cm)$)},legend columns=7,fill=none,draw=black,anchor=center,align=center},
ylabel = {F1 (in \%)},
legend to name=fredself1,
title = \walmartamazon,
mark size=1pt]
\addplot [gray,mark=triangle*] coordinates {
(512,55.54216099858763)
(640,59.43548361835075)
(768,57.87141480333698)
(896,59.213278147277705)
(1024,62.82076945998942)
(1152,60.956711101759296)
(1280,58.826191720946085)};
\addplot [blue,mark=diamond*] coordinates {
(512,70.42215981498671)
(640,72.36031420315379)
(768,74.21936963658538)
(896,75.30940835398916)
(1024,77.60824166403194)
(1152,75.8516552716273)
(1280,78.17054239336255)};
\addplot[red!70!black,mark=*] coordinates {
(512,71.23019265418745)
(640,74.47987672750745)
(768,76.69416112950653)
(896,77.31503935139213)
(1024,78.91112812797309)
(1152,80.13557951863942)
(1280,79.05895690926916)};
\addplot[brown,mark=star] coordinates {
(512,71.60215674906419)
(640,75.84674156932807)
(768,79.0403100144895)
(896,80.92390374562557)
(1024,82.95706790337137)
(1152,85.07223227108285)
(1280,85.39882140948545)};
\addplot[cyan,mark=square*] coordinates {
(512,74.6293441871921)
(640,80.94461315875243)
(768,84.06772934792744)
(896,86.31644297263504)
(1024,88.09911137379281)
(1152,89.16086380794205)
(1280,90.53106382703947)};
\addplot[green,mark=otimes*] coordinates {
(512,73.66863423318287)
(640,80.14933255808984)
(768,83.70039398829427)
(896,85.57721322181177)
(1024,88.52173792312877)
(1152,89.74831530690267)
(1280,90.71749966656927)};
\addplot[black,mark=pentagon*] coordinates {
(512,76.40145457161465)
(640,80.66966129175412)
(768,82.44761061086638)
(896,85.58658225931345)
(1024,87.39553364152685)
(1152,88.67605801992684)
(1280,89.79624981636837)};
\addlegendentry{Random}
\addlegendentry{Greedy};    
\addlegendentry{QBC};
\addlegendentry{Partition-4};
\addlegendentry{BADGE};
\addlegendentry{Partition-2};
\addlegendentry{Uncertainty};          
\coordinate (c1) at (rel axis cs:0,1);

\nextgroupplot[label style={font=\normalsize},
xtick={500, 750, 1000, 1250},
tick label style={font=\normalsize},
title = \amazongoogle,
mark size=1pt]
\addplot [gray,mark=triangle*] coordinates {
(512,59.51163003591634)
(640,60.36383886638623)
(768,61.33707597387426)
(896,61.97742181433408)
(1024,61.994958591326565)
(1152,63.31110446171003)
(1280,63.03711476863029)};
\addplot [blue,mark=diamond*] coordinates {
(512,65.32003871895836)
(640,66.5175802993319)
(768,68.01482963714363)
(896,69.87614213363183)
(1024,71.33710028656343)
(1152,73.3412826567874)
(1280,74.91482653609687)};
\addplot[red!70!black,mark=*] coordinates {
(512,63.097359003452446)
(640,66.3889366024218)
(768,67.93260257704972)
(896,69.95359894080049)
(1024,71.07758829206553)
(1152,73.09883236523655)
(1280,75.1986443348093)};
\addplot[brown,mark=star] coordinates {
(512,63.42497226955926)
(640,66.02512344671497)
(768,66.90372771330004)
(896,69.31937643874195)
(1024,70.78999186535549)
(1152,73.3076963469457)
(1280,74.49705791240294)};
\addplot[cyan,mark=square*] coordinates {
(512,68.46523644362298)
(640,70.3692742521215)
(768,74.43557650677131)
(896,77.75977425579357)
(1024,79.30940709141608)
(1152,81.48180019874293)
(1280,82.76160248112055)};
\addplot[green,mark=otimes*] coordinates {
(512,68.51094225993029)
(640,70.17505061520963)
(768,73.2925315621678)
(896,75.71901957059936)
(1024,78.50722807274367)
(1152,79.85006498458391)
(1280,82.18944191074651)};
\addplot[black,mark=pentagon*] coordinates {
(512,68.61263283173051)
(640,69.39624871180504)
(768,73.33334305857768)
(896,76.11607934125945)
(1024,78.69278923350308)
(1152,80.38467005148897)
(1280,82.0731115558397)};
\nextgroupplot[label style={font=\normalsize},
xtick={500, 750, 1000, 1250},
tick label style={font=\normalsize},
title = \dblpacm,
mark size=1pt]
\addplot [gray,mark=triangle*] coordinates {
(512,96.95447573584511)
(640,97.07809527118172)
(768,97.70201342023358)
(896,97.90808023070556)
(1024,97.91746533705168)
(1152,97.9278871273953)
(1280,97.78598588744215)};
\addplot [blue,mark=diamond*] coordinates {
(512,88.34511467898719)
(640,90.63848483814114)
(768,90.64728019541803)
(896,88.89151380638536)
(1024,89.80072601673449)
(1152,87.24480846307605)
(1280,89.98810222004771)};
\addplot[red!70!black,mark=*] coordinates {
(512,98.45110269465506)
(640,98.43715829034429)
(768,98.62467617020266)
(896,98.67520686596674)
(1024,98.8018014884008)
(1152,98.68293677838534)
(1280,98.81733212377175)};
\addplot[brown,mark=star] coordinates {
(512,98.38481138609878)
(640,98.67414488576586)
(768,98.79186577392434)
(896,98.99689523364236)
(1024,99.00392036911833)
(1152,99.05034308754327)
(1280,99.02634836950445)};
\addplot[cyan,mark=square*] coordinates {
(512,98.68472720423199)
(640,98.87067694341435)
(768,98.86428964843054)
(896,98.9935971785798)
(1024,99.03656135947044)
(1152,99.08891451125378)
(1280,99.05917307954192)};
\addplot[green,mark=otimes*] coordinates {
(512,98.79520587003341)
(640,98.8941832982294)
(768,99.03586651391724)
(896,99.01299845192213)
(1024,99.02766138464592)
(1152,99.09546164197207)
(1280,99.0948996864715)};
\addplot[black,mark=pentagon*] coordinates {
(512,98.73743526349831)
(640,98.91382835724825)
(768,98.93846998823066)
(896,99.06541587746047)
(1024,99.05805615629387)
(1152,99.12610521265577)
(1280,99.12624844989074)};
\nextgroupplot[label style={font=\normalsize},
xtick={500, 750, 1000, 1250},
tick label style={font=\normalsize},
title = \dblpscholar,
mark size=1pt]
\addplot [gray,mark=triangle*] coordinates {
(512,87.37674516044898)
(640,87.75436312438629)
(768,88.17266813804284)
(896,88.79229423237769)
(1024,88.38230026484214)
(1152,88.90513747323544)
(1280,89.12113868933943)};
\addplot [blue,mark=diamond*] coordinates {
(512,82.72452119520835)
(640,84.47635673716754)
(768,84.4501783007153)
(896,86.53948345569168)
(1024,84.99190458425768)
(1152,81.5189814627925)
(1280,77.9227894260577)};
\addplot[red!70!black,mark=*] coordinates {
(512,91.02448403957327)
(640,92.19821911625759)
(768,92.75805680638828)
(896,93.27707820183765)
(1024,93.36076898331663)
(1152,93.89119059828187)
(1280,94.48592442780033)};
\addplot[brown,mark=star] coordinates {
(512,92.49429049647262)
(640,92.7574623635712)
(768,93.43310996838072)
(896,93.65736481623597)
(1024,94.0147041651596)
(1152,94.74188464140678)
(1280,95.02643494281529)};
\addplot[cyan,mark=square*] coordinates {
(512,93.60927852265547)
(640,94.40931095706647)
(768,95.23538284030795)
(896,95.7412594013687)
(1024,96.18466818659537)
(1152,96.54435255021353)
(1280,96.7630113724318)};
\addplot[green,mark=otimes*] coordinates {
(512,93.4664956144145)
(640,94.23805556891264)
(768,94.98020904103672)
(896,95.50079424876968)
(1024,96.12377497824828)
(1152,96.52485473683386)
(1280,96.82334906058011)};
\addplot[black,mark=pentagon*] coordinates {
(512,93.5204838600612)
(640,94.29579784294152)
(768,95.11604546981508)
(896,95.56944029987827)
(1024,96.11163420073115)
(1152,96.6337401666118)
(1280,96.81335937980654)};
\nextgroupplot[label style={font=\normalsize},
xtick={500, 750, 1000, 1250},
tick label style={font=\normalsize},
title = \abtbuy,
mark size=1pt]
\addplot [gray,mark=triangle*] coordinates {
(512,66.85543251760288)
(640,74.52324021271846)
(768,73.13018350584119)
(896,72.7601201638969)
(1024,75.54502446990872)
(1152,76.18540118185516)
(1280,78.20093948439707)};
\addplot [blue,mark=diamond*] coordinates {
(512,68.48913799162965)
(640,76.45034776436007)
(768,74.81822915730973)
(896,76.20378606432722)
(1024,79.31597875777081)
(1152,79.46987393637671)
(1280,79.90430843730813)};
\addplot[red!70!black,mark=*] coordinates {
(512,76.60122611255873)
(640,78.6486859617633)
(768,79.64642782999226)
(896,81.8324902673433)
(1024,83.16932319837042)
(1152,82.53333741066632)
(1280,83.90825825917285)};
\addplot[brown,mark=star] coordinates {
(512,79.68949336740071)
(640,82.9852008505245)
(768,81.75824104696684)
(896,86.9104318150571)
(1024,88.7858241546301)
(1152,89.73029607307899)
(1280,90.63791011122512)};
\addplot[cyan,mark=square*] coordinates {
(512,83.83666709937896)
(640,86.53392086419802)
(768,89.1955944603986)
(896,90.1957755404935)
(1024,91.00426810121971)
(1152,92.27721680810926)
(1280,92.45616521989074)};
\addplot[green,mark=otimes*] coordinates {
(512,80.19825058936453)
(640,86.31020782562744)
(768,88.19215725909318)
(896,90.52144942506104)
(1024,92.3934254550513)
(1152,93.05192989041771)
(1280,93.59501337412436)};
\addplot[black,mark=pentagon*] coordinates {
(512,81.86766560458928)
(640,83.58617161494251)
(768,86.10984851888477)
(896,89.33562286900232)
(1024,90.43025593016768)
(1152,91.78851719032579)
(1280,92.31247237027583)};

\coordinate (c2) at (rel axis cs:1,1);

\end{groupplot}
\coordinate (c3) at ($(c1)!.5!(c2)$);
\node[below] at (c3 |- current bounding box.south)
{\pgfplotslegendfromname{fredself1}};

\end{tikzpicture}
\caption{Comparison of DIAL with different selection methods on All Pairs Evaluation against increasing number of instances selected by active learning}
\label{fig:selectionallf1}
\end{figure*}
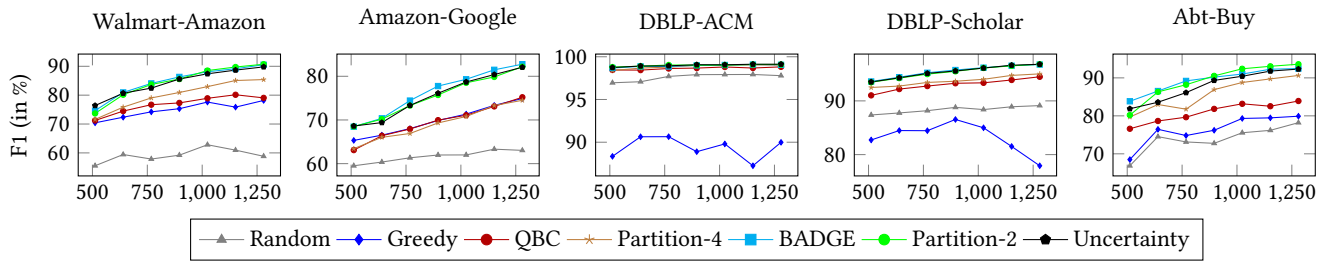
\begin{table}
    \centering
          \caption{Comparison of \sysname\ with different example selection strategies on F1 scores evaluated on all pairs after $10$ rounds of active learning. \sysname\ is agnostic to the choice of selection strategy, and hence can operate with many different methods used in the active learning literature.}
    \begin{tabular}{l|c|c|c|c|c}
        \toprule
        Method & W-A & A-G & D-A & D-S & A-B \\
        \midrule
        Random & $58.8$ & $63.0$ & $97.8$ & $89.5$ & $78.2$  \\
        Greedy & $78.2$ & $74.9$ & $90.0$ & $77.9$ & $79.9$  \\
        Partition-$2$ & $90.7$ & $82.2$ & $99.1$ & $96.8$ & $93.2$  \\
        Partition-$4$ & $85.4$ & $74.5$ & $99.0$ & $95.0$ & $90.6$  \\
        QBC & $79.1$ & $75.2$ & $98.8$ & $94.6$ & $83.9$  \\
        BADGE & $90.5$ & $82.8$ & $99.1$ & $96.8$ & $92.5$  \\
        Uncertainty & $89.8$ & $82.1$ & $99.1$ & $96.8$ & $92.3$  \\
        \bottomrule
    \end{tabular}
    \label{tab:selection}
\end{table}

Figure~\ref{fig:selectionallf1} shows All-Pair F1 at each step of active learning and 
Table $\ref{tab:selection}$ reports the All-Pair F1 scores after $10$ active learning rounds on each of the $5$ datasets using different selection strategies. We find that Partition-2 and BADGE provide gains over plain uncertainty sampling, as well as beat all other strategies by a high margin establishing their effectiveness for active learning.  

\subsection{Running time}
\label{sec:runningtime}


Table \ref{tab:alltimes} reports the time required by the different operations of \sysname\ in the $10^{th}$ round of active learning on all datasets. We emphasize here that matcher and committee training times are cumulative training times, i.e. we measure the time taken to train on all data labeled so far. We notice that the committee training time is comparable to matcher training time, despite the fact that the committee is trained for $10\mathrm{x}$ more epochs than the matcher. 
The testing time of \sysname\ with different committee sizes is reported in Table \ref{tab:alltesttimes}. 
We notice that as the committee size is increased from $\numC=1$ to $10$, the corresponding testing time increases by less than $5\%$ establishing the scalability of Index-By-Committee. 
\begin{table}
    \centering
          \caption{Time taken, in seconds, by the different operations of \sysname\ in the $10^{th}$ round of active learning}
    \begin{tabular}{l|r|r|r|r|r}
        \toprule
        Operation & W-A & A-G & D-A & D-S & A-B \\
        \midrule
        Train Matcher & $109.8$ & $71.5$ & $147.0$ & $110.1$ & $161.9$\\
        Train Committee & $102.0$ & $132.2$ & $141.2$ & $145.7$ & $35.3$\\
        Indexing \& Retrieval & $1.8$ & $0.4$ & $0.5$ & $4.8$ & $0.2$\\
        Selection & $73.0$ & $6.0$ & $8.9$ & $221.9$ & $34.71$\\
        \bottomrule
    \end{tabular}
    \label{tab:alltimes}
\end{table}
\begin{table}
    \centering
       \caption{Testing Times (in seconds) of \sysname\, with different committee sizes}
    \begin{tabular}{l|c|c|c|c|c}
        \toprule
        Method & W-A & A-G & D-A & D-S  & A-B\\
        \midrule
        \sysname\ $(\numC=1)$ & $87.6$ & $7.9$ & $15.5$ & $254.8$& $41.8$\\
        \sysname\ $(\numC=3)$ & $88.3$ & $8.0$ & $15.6$ & $256.7$ & $42.0$\\
        \sysname\ $(\numC=10)$ & $90.8$ & $8.2$ & $15.8$ & $263.1$ & $42.0$\\
        \bottomrule
    \end{tabular}
    \label{tab:alltesttimes}
    
\end{table}

\section{Related Work}
\sysname~lies in the intersection of four distinct research areas: deep learning, entity resolution, active learning, and blocking for ER. In this section, we provide further details supporting the discussion presented in \mysecref{sec:introduction} and review work from the broader literature.
\subsection{Active Learning for Entity Resolution}
\eat{\begin{itemize}
    \item Interactive Deduplication using Active Learning \cite{10.1145/775047.775087}
    \item Active Learning for Large-Scale Entity Resolution \cite{DBLP:conf/cikm/QianPS17}
    \item SystemER: A Human-in-the-loop System for Explainable Entity Resolution \cite{DBLP:journals/pvldb/QianPS19}
    \item Low-resource Deep Entity Resolution with Transfer and Active Learning  \cite{DBLP:conf/acl/KasaiQGLP19}
    \item A Comprehensive Benchmark Framework for Active Learning Methods in Entity Matching \cite{DBLP:conf/sigmod/Meduri0SS20}
    \item Learning-Based Methods with Human-in-the-Loop for Entity Resolution \cite{DBLP:conf/cikm/Gurajada0QS19}
    \item Unsupervised Bootstrapping of Active Learning for Entity Resolution \cite{10.1007/978-3-030-49461-2_13}
    \item Heterogeneous Committee-Based Active Learning for Entity Resolution (HeALER) \cite{10.1007/978-3-030-28730-6_5}
\end{itemize}}
Over the years, a number of works have applied active learning to ER using a variety of (paired) classifiers including support vector machines, decision trees \citep{sarawagi02,tejada2001learning}, explainable ER rules \citep{10.1145/1807167.1807252,DBLP:conf/cikm/QianPS17,DBLP:journals/pvldb/QianPS19}. However, most of these assume that the blocking function is known. In fact, some of the aforementioned works that attempt to learn rules \citep{DBLP:conf/cikm/QianPS17,DBLP:journals/pvldb/QianPS19} ask the user to mark every possible blocker in the input feature space. DIAL attempts to improve upon these approaches by not only \emph{learning} the blocker but also via the use of a more powerful paired classifier (pre-trained language models). \citet{DBLP:conf/sigmod/Meduri0SS20} provide an in-depth comparison of various matchers and example selectors but neither consider TPLMs nor address how to learn a blocker. While HEALER \citep{10.1007/978-3-030-28730-6_5} attempts to improve upon \citet{10.14778/2735471.2735474}'s committee-based approach to ER by including different kinds of matchers, it does not consider neural networks or TPLMs in its heterogeneous committee.

\par

Alongside AL for ER, another line of work attempts to solve ER by crowd-sourcing \citep{10.14778/2350229.2350263}. The drawbacks of this approach include that no model is learned (neither matcher nor blocker) thus incurring monetory costs needed to pay the crowd each time we are faced with new data to deduplicate. Distinct from DIAL's, their focus is more towards correcting the labels obtained from the crowd (may not constitute experts) \citep{10.1145/3183713.3183755} and most efficient interface to elicit most labels at least cost \citep{10.1145/3035918.3035931}.

\subsection{Deep Learning for Entity Resolution}
Deep Learning has been used to tackle various aspects of the Entity Resolution task including blocking \cite{10.14778/3236187.3236198, 10.1145/3336191.3371813}, and matching \cite{deepmatcher, DBLP:conf/edbt/BrunnerS20, ditto, 10.1145/3308558.3313578, tracz-etal-2020-bert, 9338287}, and detecting variations \cite{embar2020contrastive} which are duplicates on a given set of base attributes but differ on other attributes. The example used in the Section \ref{sec:paired} of a pair of records describing two different editions of the same book is an example of a variation. We refer the reader to \cite{barlaug2020neural} for an extensive survey on deep learning for entity matching. Deep learning methods for ER can be broadly classified into methods which operate on separate embeddings of instances $r$ and $s$ \cite{10.14778/3236187.3236198, 10.1145/3336191.3371813, ditto}, and those which operate on the joint embedding of the record pair $(r, s)$ \cite{deepmatcher, ditto, DBLP:conf/edbt/BrunnerS20, 10.1145/3357384.3358018, tracz-etal-2020-bert}. While joint embeddings provide more information useful for ER, DIAL shows there is a place for both, i.e., jointly embedding the pair and embedding them separately for use in blocking. DeepMatcher \citep{deepmatcher} proposed one of the first neural network architectures for ER, which was improved upon by DITTO \citep{ditto}. Neither of these consider blocking nor active learning. In an effort to tackle ER in low-resource settings such as scarcely available labeled data, DTAL \citep{DBLP:conf/acl/KasaiQGLP19} proposes learning a neural network via active learning with uncertainty sampling along with partitioning but does not consider TPLMs and neither addresses learning a blocker. \sysname\ improves upon DTAL by learning an integrated matcher and blocker where the matcher is a more powerful TPLM, and DITTO’s advanced blocking in the
active learning setting as shown via our experiments.
\subsection{Active Learning for Deep Learning}
Perhaps the closest work to our setting is \cite{mussmann-etal-2020-importance}, which also considers TPLMs for active learning on pairwise classification tasks. They use a similar architecture as \cite{gillick-etal-2019-learning}, i.e. a TPLM invoked and trained in single mode to retrieve similar embeddings. Key differences from \sysname\ are 1) They do not use random negatives, 2) They do not consider separate models for matching and blocking, 3) They do not create a committee of multiple embeddings. 

At the intersection of committee based methods for active learning, and deep learning, lies \cite{8579074} which creates a committee of Convolutional Neural Networks (CNN) based classifiers for Active Learning in Image Classification. The area of Deep Active Learning is rapidly growing with exciting works like BALD \cite{gal2017deep}, Loss Prediction \cite{yoo2019learning}, and Batch Aware methods like BatchBALD \cite{kirsch2019batchbald}, BADGE \cite{badge} and \cite{sener2018active, zhdanov2019diverse}. A comprehensive survey of deep active learning methods can be found in \cite{ren2020survey}. As stated earlier, most of these are compatible for use as example selectors in DIAL.

\eat{\begin{itemize}
    \item On the Importance of Adaptive Data Collection for Extremely Imbalanced Pairwise Tasks \cite{mussmann-etal-2020-importance}
    \item The Power of Ensembles for Active Learning in Image Classification \citep{8579074}: Not about paired classification but definitely builds a committee of CNNs
\end{itemize}}

\subsection{Blocking}
Besides hand-coded blocking functions, earlier methods for blocking relied on unsupervised clustering \citep{10.1145/347090.347123} and passive learning with labeled data required up-front \citep{bilenko2006adaptive}. The latter uses red-blue set cover to learn an effective blocking function but its need for labeled data makes it ineffective in settings that call for active learning. While other approaches for blocking are available, a number of these utilize unsupervised learning~\citep{blockedmatmul,10.1007/978-3-030-49461-2_13,Galhotra_2021}. 

Token Blocking~\citep{6255742} uses tokens from every attribute value as blocking keys, and records with common tokens are put in one block. This yield high recall at the cost of low precision. Several methods have been proposed to deal with redundant and superfluous comparisons~\citep{6255742,10.14778/2856318.2856326,10.14778/2733085.2733098,5887335,1552993,10.1145/223784.223807,10.1145/347090.347123}.
Meta-Blocking~\citep{6487505} operates on redundancy-positive block collections where the number of shared blocks indicate likelihood of matching. A blocking graph is created from the given redundancy-positive block collection, and is pruned using matching likelihoods~\citep{6487505,10.14778/2994509.2994533,Zhang_2017,DBLP:conf/edbt/0001PPK16,DBLP:conf/edbt/Efthymiou0SC19,10.14778/2733085.2733098,DALBIANCO201875}. 

AutoBlock \cite{10.1145/3336191.3371813} assumes knowledge of strong attributes (e.g., UPC code for grocery products), that may be used to produce labeled data for learning the blocking function. DIAL does not make any such assumptions and can work with heterogeneous lists. Both AutoBlock and DeepER \cite{10.14778/3236187.3236198} use Locality Sensitive Hashing (LSH) for retrieval, and DITTO uses similarity search by blocked matrix multiplication \cite{blockedmatmul}. In contrast, \sysname\ uses FAISS \cite{faiss}, a highly optimized library for k-selection which relies on product quantization for fast asymmetric distance computations. 

Another related task is training a retrieval system for entity linkage~\cite{gillick-etal-2019-learning}. Key similarities with our blocker model include fine-tuning the TPLM in the single mode, and using random negatives to train the TPLM. This work differs from our work in that it does not perform active learning. 
We refer the reader to~\citep{10.1145/3377455,KOPCKE2010197,5887335,10.1145/3418896} for an extensive survey on blocking. 

\section{Conclusion}
In this work we present \sysname, a scalable active learning system with an integrated matcher-blocker combination. As opposed to most works in ER, DIAL~ learns the blocker in addition to the matcher. Furthermore, the blocker and matcher are integrated in a way so that improvements in one can benefit the other. We show that our approach leads to improved recall during blocking and improved matching via the use of transformer-based pre-trained language models. We successfully train a committee on top of powerful TPLM representations, and use it to perform Index-by-Committee, a novel and efficient example retrieval technique. Our experimental results on $5$ real world datasets show that \sysname\ outperforms baseline methods by a large margin while also requiring minimal human involvement. We showcase our approach by reporting the effectiveness of \sysname\ on a multilingual dataset where hand-coding a blocking function may not be possible due to the different languages involved.  

{
\bibliographystyle{ACM-Reference-Format}
\bibliography{sample}
}
\end{document}